\documentclass[a4paper,11pt]{article}
\pdfoutput=1
\usepackage{jheppub}

\usepackage{epsf}
\usepackage{epsfig}
\usepackage{subfigure}
\usepackage{mathtools}
\usepackage{hhline}
\usepackage{float}
\usepackage{multirow}
\usepackage{nicefrac}
\usepackage{epstopdf}
\usepackage{slashed}
\usepackage{xcolor}
\usepackage{url}
\usepackage{umoline}

\newcommand{\dis}[1]{\begin{equation}\begin{split}#1\end{split}\end{equation}}

\newcommand{\be}{\begin{eqnarray}}
\newcommand{\ee}{\end{eqnarray}}

\newcommand{\ba}{\begin{array}}
\newcommand{\ea}{\end{array}}
\newcommand{\bee}{\begin{equation}\ba{c}}
\newcommand{\eee}{\ea\end{equation}}

\newcommand{\bi}{\begin{itemize}}
\newcommand{\ei}{\end{itemize}}

\toccontinuoustrue

\title{Cascade decays of Heavy Higgs bosons through vectorlike quarks}
\author{Radovan Dermisek$^1$,}
\author{Enrico Lunghi$^1$}
\author{and Seodong Shin$^{2,3,4,5}$}
\affiliation{
$^1$Physics Department, Indiana University, Bloomington, IN 47405, USA \\
$^2$Enrico Fermi Institute, University of Chicago, Chicago, IL 60637, USA \\
$^3$Department of Physics and IPAP, Yonsei University, Seoul 03722, Korea \\
$^4$Center for Theoretical Physics of the Universe, Institute for Basic Science, Daejeon 34126, Korea\\
$^5$Department of Physics, Jeonbuk National University, Jeonju, Jeonbuk 54896, Korea \\
}
\emailAdd{dermisek@indiana.edu}
\emailAdd{elunghi@indiana.edu}
\emailAdd{sshin@jbnu.ac.kr}

\abstract{
We study cascade decays of heavy neutral Higgs bosons through vectorlike quarks. We focus on scenarios where decay modes into pairs of vectorlike quarks are not kinematically open which extends the sensitivity of the LHC to larger masses. Assuming only mixing with the third family of standard model quarks the new decay modes of  heavy Higgs bosons are:  $H\to t_4 t \to Wbt, Zt t, ht t$ and $H\to b_4 b \to Wtb, Zb b, hb b$, where $t_4$ ($b_4$) is the new up-type (down-type) quark mass eigenstate. In the numerical analysis we assume the CP even Higgs boson in the two Higgs doublet model type-II but the signatures are relevant for many other scenarios. We identify the region of the parameter space where these decay modes are significant or can even dominate, and thus they provide the best opportunities for the simultaneous discovery of a new Higgs boson and vectorlike quarks. We further explore the reach of the High Luminosity LHC for two representative decay modes, $t_4\to Zt\to\ell\ell t$ and $b_4\to Zb\to\ell\ell b$, and found that cross sections at a 0.1 fb level can be probed with simple cut based analyses. We also find that the rates for Higgs cascade decays can be much larger than the rates for a single production of vectorlike quarks. Furthermore, the reach for vectorlike quarks in Higgs cascade decays and pair production extends to comparable masses. \newpage
}

\begin{document}

\maketitle

\section{Introduction}
\label{sec:introduction}

Among the simplest extensions of the standard model (SM) are models with extra Higgs bosons or vectorlike matter. Many searches for such  individual new particles have been performed. However, if both sectors are present, new search strategies can be designed that could lead to a simultaneous discovery of heavy Higgs bosons and matter particles. These can be even more potent than separate searches. In this paper we focus on cascade decays of a heavy neutral Higgs boson through  vectorlike quarks.

\begin{figure}
\begin{center}
\includegraphics[width=.7\linewidth]{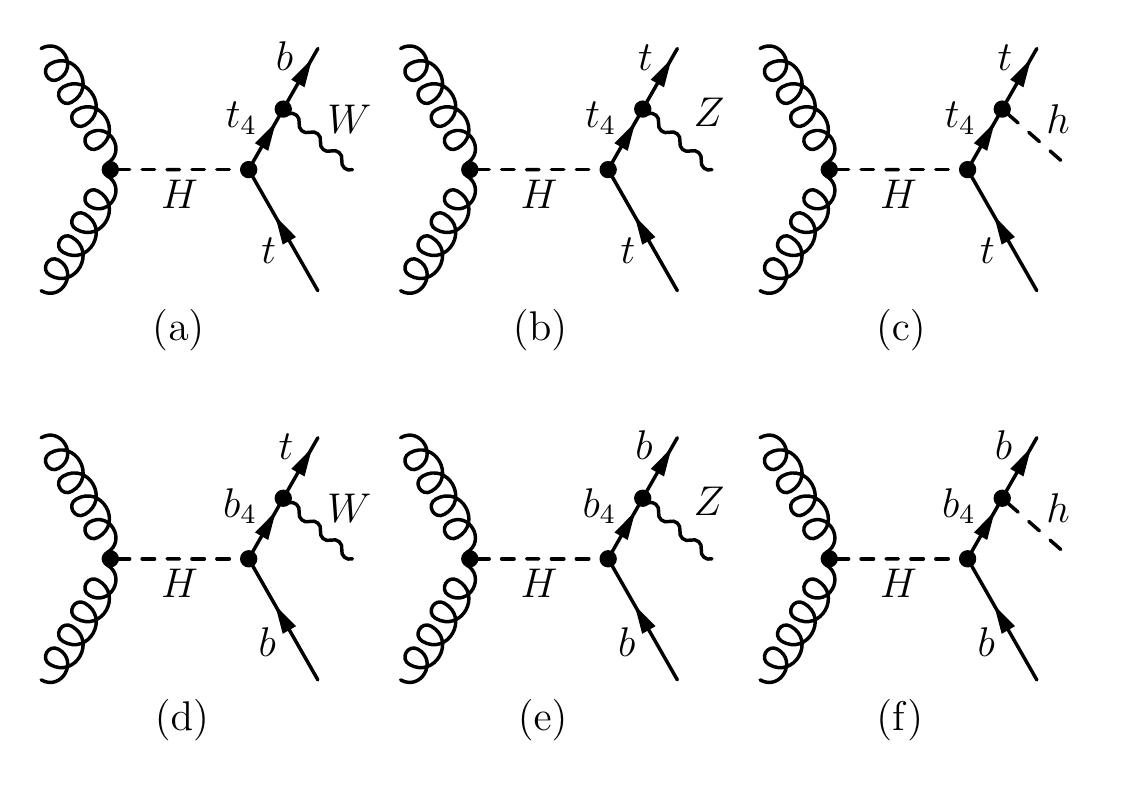}
\caption{New decay topologies of a heavy Higgs boson through vectorlike quarks.}
\label{fig:topologies}
\end{center}
\end{figure}

In this work, we consider an extension of the two Higgs doublet model type-II by vectorlike pairs of new quarks (VLQ), corresponding to a copy of the  SM SU(2) doublet and singlet quarks and their vectorlike partners, introduced in ref.~\cite{Dermisek:2019vkc}.  Assuming only the mixing with the third family of standard model quarks, the flavor changing couplings of  $W$, $Z$ and Higgs bosons between new quarks and the third family quarks are generated. These couplings allow new decay modes of the heavy CP even (or CP odd) Higgs boson: $H \to t_4 t$ and $H \to b_4 b$, where $t_4$ and $b_4$ are the  lightest  new up-type and down-type quark mass eigenstates.  Although these decay modes compete with $H \to t \bar t$ and $H \to b \bar b$ we will see that,  in a region of the parameter space, they are significant or can even dominate. Here we assume that the light Higgs boson ($h$) is SM-like so that $H \to ZZ,\; WW$ are  not present. Subsequent decay modes of $t_4$ and $b_4$: $t_4 \to W  b$, $t_4 \to Z t$,  $t_4 \to h t$ and $b_4 \to W t$, $b_4 \to Z b$, $b_4 \to h b$ lead to the following 6 decay chains of the heavy Higgs boson:
\begin{align}
H &\; \to \; t_4 t \; \to \; Wbt, \;  Zt t,\; h t t, \label{eq:Ht4} \\
H &\; \to \; b_4 b \; \to \; Wtb, \; Zb b,\; h b b, \label{eq:Hb4}
\end{align}
 which are also depicted in figure~\ref{fig:topologies}.

In addition, a heavy neutral Higgs boson could also decay into pairs of vectorlike quarks. However, we focus on the range of masses where  $H \to t_4 \bar t_4, \; b_4 \bar b_4$ are not kinematically open which extends the sensitivity of the LHC to larger masses of vectorlike quarks (moreover, the final states would be the same as for the pair production of vectorlike quarks). We also find that the rates for processes  (\ref{eq:Ht4}) and (\ref{eq:Hb4}) can be much larger than the rates for a single production of vectorlike quarks. Thus, in the identified regions, they provide the best opportunities for the discovery of a new Higgs boson and vectorlike quarks.  Although in the numerical analysis we assume the heavy CP even Higgs boson in the two Higgs doublet model type-II, the signatures are relevant for many other scenarios. 

Similar signatures in the lepton sector, cascade decays of heavy Higgs bosons through vectorlike leptons, were studied in refs.~\cite{Dermisek:2015oja, Dermisek:2015vra, Dermisek:2015hue, Dermisek:2016via, CidVidal:2018eel}. If decay modes through both vectorlike quarks and leptons are kinematically open, the decays through quarks are expected to dominate because of the color factor. On the other hand, decay modes through leptons provide several very clean signatures~\cite{Dermisek:2015hue, Dermisek:2014qca} that might compensate for smaller rates.
Alternatively, in the same model, if vectorlike quarks or leptons are heavier, they can decay through heavy Higgs bosons, including the charged Higgs boson, leading to very rare final states.
The corresponding signatures were recently studied in ref.~\cite{Dermisek:2019vkc}.

Vectorlike quarks and leptons near the electroweak scale provide a very rich phenomenology and often they are introduced to explain various anomalies. Examples include discrepancies in precision Z-pole observables~\cite{Choudhury:2001hs, Dermisek:2011xu, Dermisek:2012qx, Batell:2012ca} and the muon g-2 anomaly~\cite{Kannike:2011ng, Dermisek:2013gta} among many others. They are also considered for a variety of theoretical reasons. Examples  include studies of their effects on gauge and Yukawa couplings in the framework of grand unification~\cite{Babu:1996zv, Kolda:1996ea, Ghilencea:1997yr, AmelinoCamelia:1998tm, BasteroGil:1999dx, Dermisek:2012as, Dermisek:2012ke, Dermisek:2017ihj, Dermisek:2018hxq}, on electroweak symmetry breaking~\cite{Dermisek:2016tzw}, and the Higgs boson mass and its decays~\cite{Moroi:1991mg, Moroi:1992zk, Babu:2008ge, Martin:2009bg, Graham:2009gy, Martin:2010dc, Falkowski:2014ffa, Dermisek:2014cia}. The supersymmetric extension with a complete vectorlike family also provides a possibility to understand the values of all large couplings in the SM  from the IR fixed point structure of the renormalization group equations~\cite{Dermisek:2018ujw}. The possibility of a simple embedding into grand unified theories is one of the main motivations for considering complete copies of SM families. We should note, however, that vectorlike quarks with other quantum numbers have been considered in the literature (see, for instance, ref.~\cite{Aguilar-Saavedra:2013qpa}). Vectorlike fermions have also been studied in composite Higgs models~\cite{Dobrescu:1997nm, Chivukula:1998wd, He:2001fz, Hill:2002ap, Agashe:2004rs, Contino:2006qr, Barbieri:2007bh, Anastasiou:2009rv,Cacciapaglia:2019ixa, Cai:2018tet, Ma:2015gra, Serra:2015xfa, Bertuzzo:2012ya, Mrazek:2011iu}, little Higgs models~\cite{ArkaniHamed:2002qy, Schmaltz:2005ky} and models with a gauged flavor group~\cite{Davidson:1987tr, Babu:1989rb, Grinstein:2010ve, Guadagnoli:2011id}. Most of these scenarios focus on mixing between the new vectorlike fermions and the third generation of SM quarks and lead to some generic signatures (see, for instance, refs.~\cite{AguilarSaavedra:2009es,Okada:2012gy,DeSimone:2012fs,Buchkremer:2013bha,Aguilar-Saavedra:2013qpa}).

This paper is organized as follows. In section~\ref{sec:model} we briefly summarize the model. Details of the analysis and experimental constraints are discussed in section~\ref{sec:scan}. The main results are presented in section~\ref{sec:Hproduction}. We briefly discuss the search strategies in section~\ref{sec:Searches} and conclude in section~\ref{sec:conclusions}. The appendix contains  formulas for partial decay widths of the heavy Higgs boson.

\section{Model}
\label{sec:model}
We consider an extension of the two Higgs doublet model type-II by vectorlike pairs of new quarks: SU(2) doublets $Q_{L,R}$ and SU(2) singlets $T_{L,R}$ and $B_{L,R}$. The $Q_L$, $T_R$ and $B_R$ have the same quantum numbers as the SM quark doublet $q_L$ and the right-handed quark singlets $u_R$ and $d_R$, respectively. The quantum numbers of new quarks, SM quarks and two Higgs doublets, are summarized in table~\ref{table:fieldcontents}.  The model is described in detail in ref.~\cite{Dermisek:2019vkc} and thus we just briefly summarize it here.

\begin{table}[htp]
\begin{center}
\begin{tabular}{c c c c c c c c c}
\hline
\hline
 & ~~$q_L$ & ~~$t_R$ & ~~$d_R$ & ~~$Q_{L,R}$ & ~~$T_{L,R}$ & ~~$B_{L,R}$ & ~~$H_d$ & ~~ $H_u$\\
\hline
SU(2)$_{\rm L}$ & ~~\bf 2 & ~~\bf 1 & ~~\bf 1 & ~~\bf 2 & ~~\bf 1 & ~~\bf 1 & ~~\bf 2 & ~~\bf 2 \\
U(1)$_{\rm Y}$ & ~~$\frac16$ & ~~$\frac23$ & ~~-$\frac13$ & ~~$\frac16$ & ~~$\frac23$ & ~~-$\frac13$ & ~~$\frac12$ & ~~-$\frac12$ \\
Z$_2$ & ~~+ & ~~+ & ~~-- & ~~+ & ~~+ & ~~-- & ~~-- & ~~+ \\
\hline
\hline
\end{tabular}
\end{center}
\caption{Quantum numbers of the 3rd generation standard model quarks ($q_L, t_R, d_R$), extra vectorlike quarks and the two Higgs doublets. The electric charge is given by $Q = T_3 +Y$, where $T_3$ is the weak isospin, which is +1/2 for the first component of a doublet and -1/2 for the second component. }
\label{table:fieldcontents}
\end{table}

As is characteristic for  the two Higgs doublet model type-II, we assume that the down sector couples to $H_d$ and the up sector couples to $H_u$. This can be achieved by the $Z_2$ symmetry specified in table \ref{table:fieldcontents}. The generalization to the whole vectorlike family of new fermions, including the lepton sector introduced in ref.~\cite{Dermisek:2015oja}, is straightforward. We further assume that, in the basis in which the SM quark Yukawa couplings are diagonal, the new quarks mix only with one family of SM quarks and we consider the mixing with the third family as an example. An arbitrary mixing could be easily accommodated.

The most general renormalizable Lagrangian consistent with our assumptions contains the following Yukawa and mass terms for the SM and vectorlike quarks:
\dis{
{\cal L} \supset \;  & - y_b \bar q_L  d_R H_d - \lambda_B \bar q_L  B_R H_d  -  \lambda_Q \bar Q_L d_R H_d -  \lambda \bar Q_L  B_R H_d - \bar \lambda H_d^\dagger \bar B_L  Q_R  \\
& - y_t \bar q_L  t_R H_u - \kappa_T  \bar q_L  T_R H_u - \kappa_Q  \bar Q_L  u_R H_u - \kappa  \bar{Q}_L  T_R H_u - \bar \kappa H_u^\dagger \bar{T}_L Q_R  \\
&  - M_Q \bar Q_L Q_R - M_T \bar T_L T_R - M_B \bar B_L B_R + {\rm h.c.}~,
\label{eq:lagrangian}
}
where the first term is the bottom Yukawa coupling, followed by Yukawa couplings of vectorlike quarks to $H_d$ (denoted by various $\lambda$s), the top Yukawa coupling, Yukawa couplings of vectorlike quarks to $H_u$ (denoted by various $\kappa$s), and finally by mass terms for vectorlike quarks. Note that the explicit mass terms mixing SM and vectorlike quarks, $M_{q} \bar q_L Q_R$, $M_{t} \bar T_L t_R$ and $M_{b} \bar B_L b_R$, can be removed by redefinitions of $Q_L$, $T_R$, $B_R$ and the Yukawa couplings. The components of doublets are labeled as follows:
\dis{
q_L  = \left(
\begin{array}{c}
t_L \\
b_L
\end{array}
\right),~
Q_{L,R}  = \left(
\begin{array}{c}
T_{L,R}^Q \\
B_{L,R}^Q
\end{array}
\right),~
H_d  = \left(
\begin{array}{c}
H_d^+ \\
H_d^0
\end{array}
\right),~
H_u  = \left(
\begin{array}{c}
H_u^0\\
H_u^-
\end{array}
\right).
}
We assume that the neutral Higgs components develop real and positive vacuum expectation values, $\left< H_u^0 \right> = v_u$ and $\left< H_d^0 \right> = v_d$, as in the $CP$ conserving two Higgs doublet model with $\sqrt{v_u^2 + v_d^2} = v = 174$ GeV and we define $\tan \beta \equiv v_u / v_d$. Plugging the vacuum expectation values to the Lagrangian, we obtain the mass matrices describing the mixing between the third generation and  vectorlike quarks:
\begin{align}
\left( \begin{array}{ccc} \bar t_L & \bar T_L^Q & \bar T_L \end{array}\right)
M_t
\left( \begin{array}{c}t_R  \\T_R^Q \\T_R\end{array} \right)
&=
\left( \begin{array}{ccc}\bar t_L & \bar T_L^Q& \bar T_L\end{array}\right)
\left( \begin{array}{ccc}
y_t v_u & 0 & \kappa_T v_u \\
\kappa_Q  v_u & M_Q & \kappa v_u \\
0 & \bar \kappa v_u & M_T \\
\end{array}\right)
\left(\begin{array}{c}t_R  \\T_R^Q \\T_R\end{array}\right)~,
\label{eq:mmu} \\
\nonumber \\
\left( \begin{array}{ccc}\bar b_L & \bar B_L^Q & \bar B_L\end{array}\right)
M_b
\left( \begin{array}{c}b_R  \\B_R^Q \\B_R\end{array}\right)
&=
\left( \begin{array}{ccc}\bar b_L & \bar B_L^Q & \bar B_L\end{array}\right)
\left( \begin{array}{ccc}
y_b v_d & 0 & \lambda_B v_d \\
\lambda_Q v_d & M_Q & \lambda v_d \\
0 & \bar \lambda v_d & M_B \\
\end{array}\right)
\left(\begin{array}{c}b_R  \\B_R^Q \\B_R\end{array}\right)~.
\label{eq:mmd}
\end{align}

\noindent
We label the resulting mass eigenstates as $t_i$ and $b_i$ with $i=3,4,5$, where $t_3$ and $b_3$ represent the top quark and the bottom quark.
A complete discussion of mass eigenstates, their couplings to the $W$, $Z$, and Higgs bosons, various approximate formulas, and other details  can be found in ref.~\cite{Dermisek:2019vkc}. Formulas for  partial decay widths of the heavy Higgs boson are summarized in the appendix. In the following section, we perform the numerical analysis assuming the alignment limit in which the light Higgs coupling to gauge bosons are identical to those in the SM. This is most easily, but not necessarily, attained in the decoupling limit where all the non-SM Higgs bosons are heavy~\cite{Haber:1994mt, Gunion:2002zf, Bernon:2015qea}.

\section{Parameter space scan and experimental constraints}
\label{sec:scan}
In the numerical study we scan the parameters of the model in the following ranges:
\begin{align}
M_{Q,T,B} & \in [900, 4000] \; {\rm GeV} \; , \label{scan-M}\\
\kappa_T, \kappa_Q, \kappa, \bar \kappa & \in [-1.0, 1.0] \; , \label{rangek}\\
\lambda_B, \lambda_Q, \lambda, \bar \lambda &\in [-1.0, 1.0] \; , \label{rangel}\\
\tan\beta & \in [0.3, 50] \; , \label{scan-tb}
\end{align}
and we will also comment on the impact of lowering the upper ranges in Eqs.~(\ref{rangek}) and (\ref{rangel}). Note that the signs of three Yukawa couplings are not physical and can be absorbed into a redefinition of the three vectorlike quark fields (any set of three couplings not containing both $\lambda_Q$ and $\kappa_Q$ can be chosen to be positive). For each choice of the parameters in eqs. (\ref{scan-M}-\ref{scan-tb}), $y_t$ and $y_b$ are determined iteratively to reproduce the top and bottom masses.

Due to current constraints on vectorlike quark masses, all indirect constraints have a minimal impact on our scan. We impose experimental constraints from oblique corrections and $R_b$~\cite{Tanabashi:2018oca} following the results presented in Sec.~II of ref.~\cite{Lavoura:1992np} and Sec.~IIIB of ref.~\cite{Chen:2017hak} (see also refs.~\cite{Burgess:1993vc, Anastasiou:2009rv}). We further include bounds from $h \to (\gamma \gamma, 4 \ell)$~\cite{Khachatryan:2016yec, Aaboud:2018xdt, ATLAS:2018doi} (note that in the calculation of the effective $hgg$ coupling we include effects of vectorlike quarks, that can be found, for instance, in ref.~\cite{Djouadi:2005gj, Bizot:2015zaa}), and direct searches for pair production of vectorlike quarks at the LHC~\cite{Aaboud:2018pii,hepdata:vlqpair} (for each scan point we calculate the $t_4$ and $b_4$ branching ratios into $W$, $Z$ and $h$ and compare with the constraint). We do not use searches for single production of VLQ~\cite{Sirunyan:2017tfc,ATLAS:2018qxs} since the constraints are not stronger than those from the pair production. We finally impose searches for heavy Higgs bosons: $H \to \tau^+ \tau^-$~\cite{Aaboud:2017sjh, hepdata:Htata}, $H \to \gamma \gamma$~\cite{Aaboud:2017yyg, hepdata:Hdiph}. In the alignment limit, the width $H\to h h$ is given in eq.~\ref{eq:Hhh} which impacts the $H\to h h$ branching ratio at the $O(0.1\%)$ level; hence $H\to h h$ constraints~\cite{Aaboud:2018knk} are negligible.

\section{Higgs production cross section and decays}
\label{sec:Hproduction}
Let us start by discussing the heavy neutral CP--even Higgs boson production cross section. Note that the results for the case of a CP-odd Higgs ($A$) would be very similar. In fact, the $bbH$ and $bbA$ production cross sections, which dominate at large $\tan\beta$, are essentially identical, and the gluon fusion production of $A$ is only slightly larger than the $H$ one for large Higgs masses (see, for instance, figure 3.36 of ref.~\cite{Djouadi:2005gj}). Also, in the alignment limit, the $H$ and $A$ branching ratios are almost identical.

\begin{figure}
\begin{center}
\includegraphics[width=.49\linewidth]{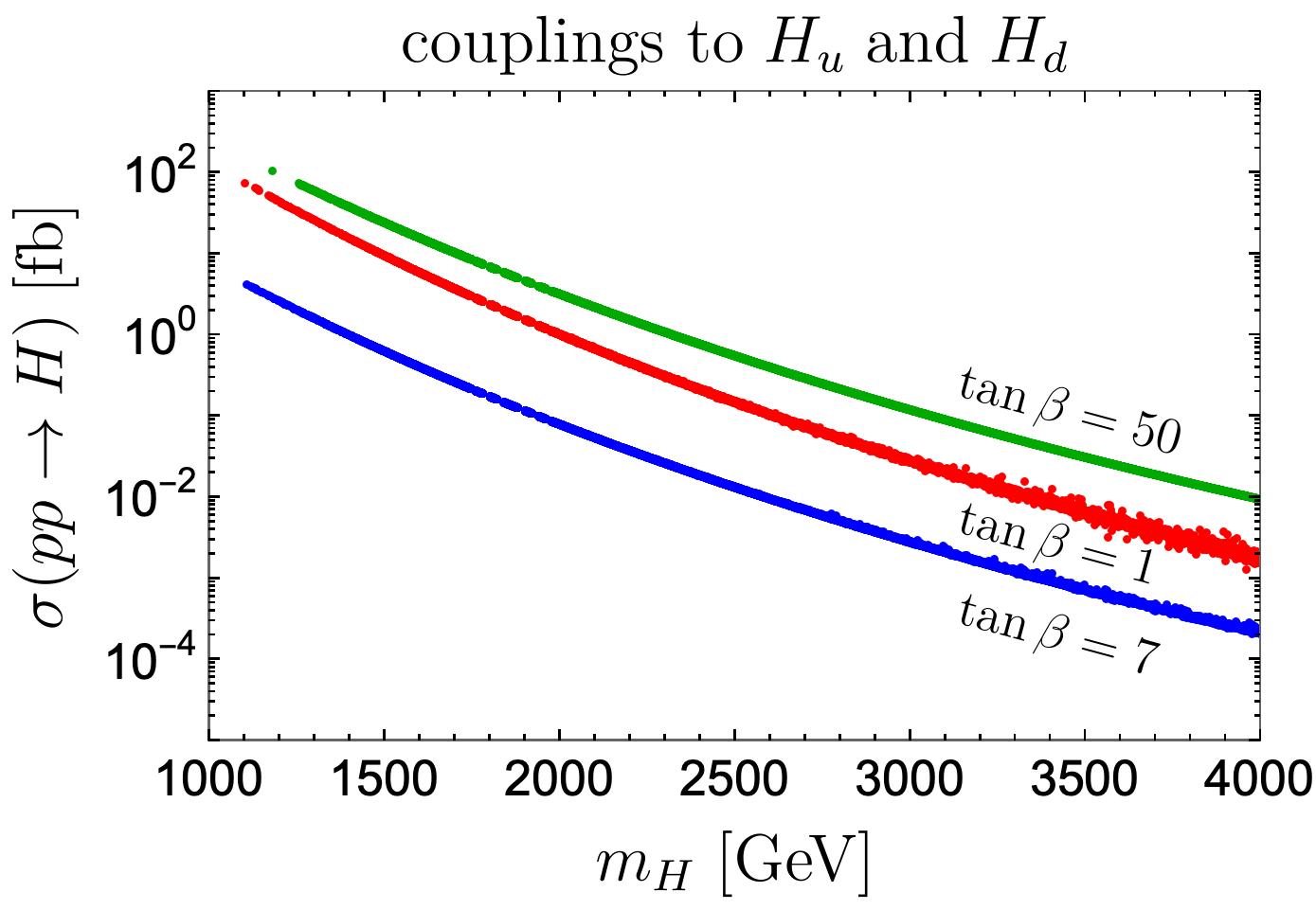}
\includegraphics[width=.49\linewidth]{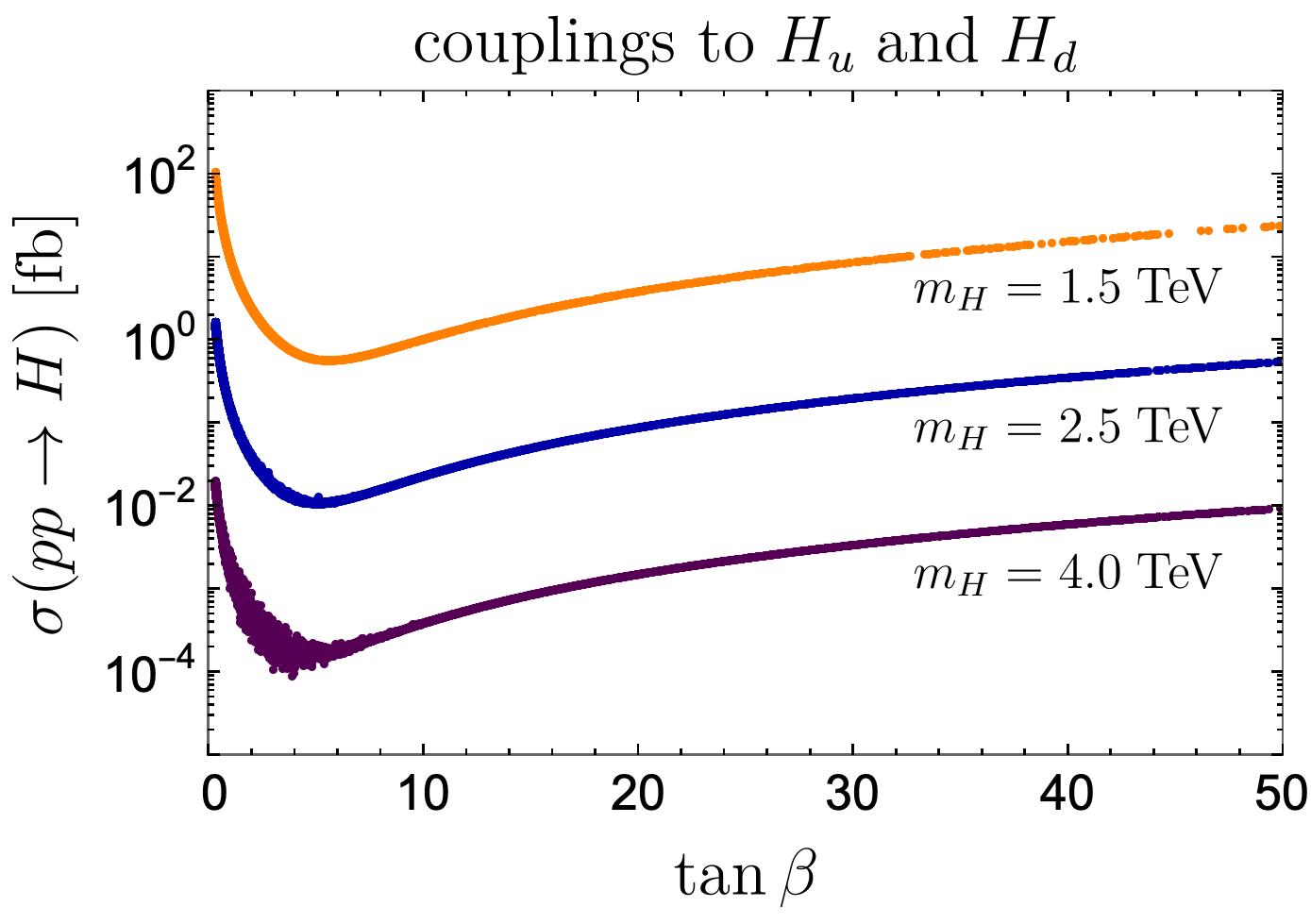}
\caption{Production cross section of a heavy neutral Higgs boson in scenarios in which decays through a vectorlike quark are open as a function of $m_H$ for $\tan \beta = 1, 7$ and 50 (left) and as a function of $\tan\beta$ for $m_H = 1.5, 2.5$ and 4 TeV (right). }
\label{fig:Hprod}
\end{center}
\end{figure}

In figure~\ref{fig:Hprod}, we show the production cross section dependence on the Higgs mass and $\tan\beta$ for scenarios in which the $H\to (t_4 t, b_4 b)$ decays are kinematically open and satisfy all experimental constraints. The lower bound on the Higgs mass in the left panel of figure~\ref{fig:Hprod} is thus connected to the limits from direct searches for vectorlike quarks. This scenario is also constrained by $H\to \tau^+ \tau^-$ searches that, for large $\tan\beta$, extend to larger Higgs masses.

The effective $ggH$ vertex is dominated by top and bottom loops, which give contributions almost identical to the type-II two Higgs doublet model ones. Vectorlike quark loops generate the spread at large Higgs mass and small $\tan\beta$. The lower bounds on vectorlike quark masses imply that the $ggH$ vertex is significantly affected only for very heavy Higgs masses (in general one expects the largest effects for $m_H \sim 2 m_{t_4, b_4}$, see, for instance, ref.~\cite{Bizot:2015zaa}). To understand the $\tan\beta$ dependence, we note that $\lambda^H_{t_4 t_4} \propto v_u \cos\beta \propto \sin 2\beta $ and that $\lambda^H_{b_4 b_4} \propto v_d \sin\beta \propto \sin 2\beta$, implying that both are the largest at $\tan\beta \sim 1$. The actual cross sections are calculated using SusHi~\cite{Harlander:2012pb} and then rescaled to take into account the impact of the modified $ggH$ vertex.

\begin{figure}
\begin{center}
\includegraphics[width=.49\linewidth]{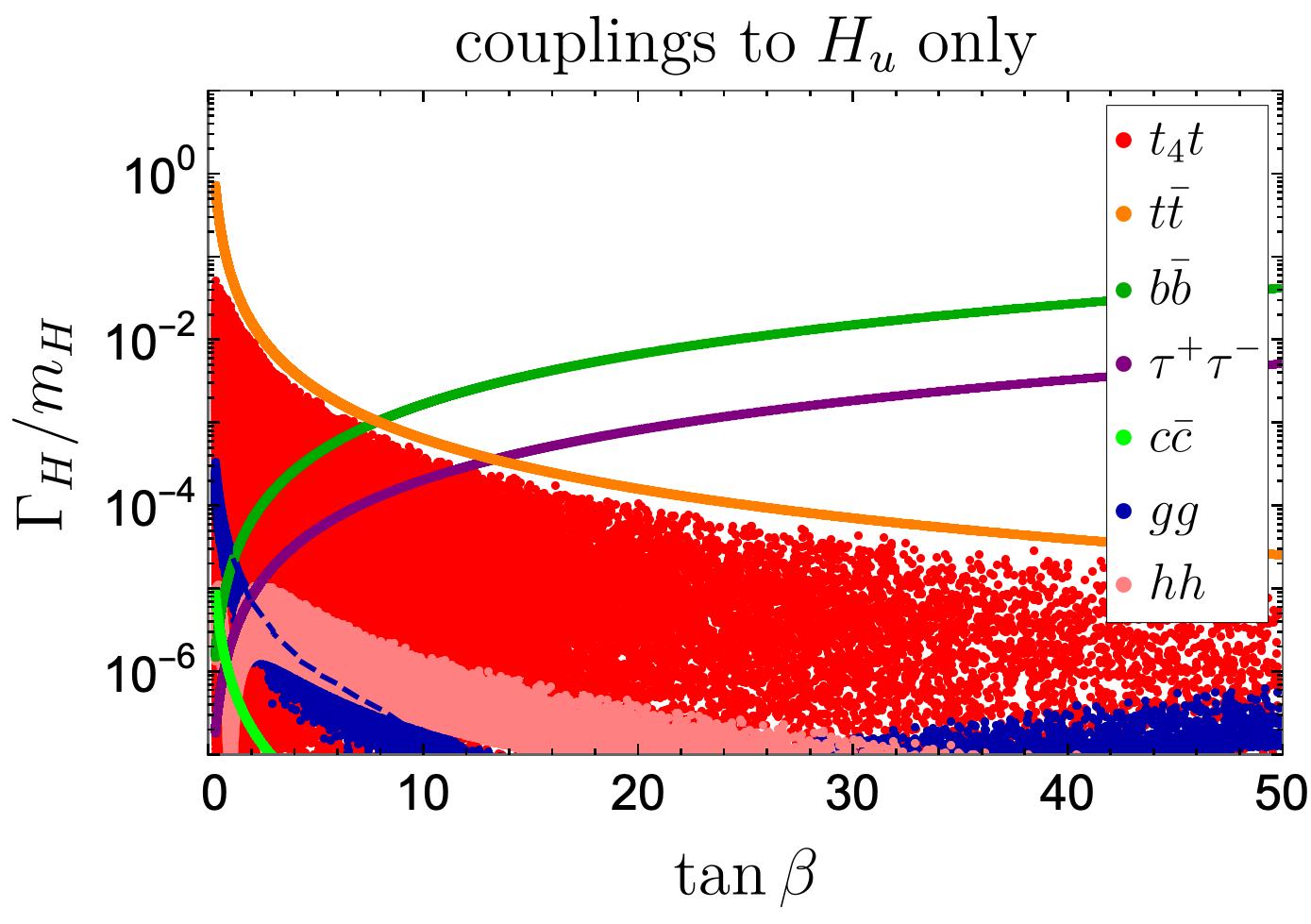}
\includegraphics[width=.49\linewidth]{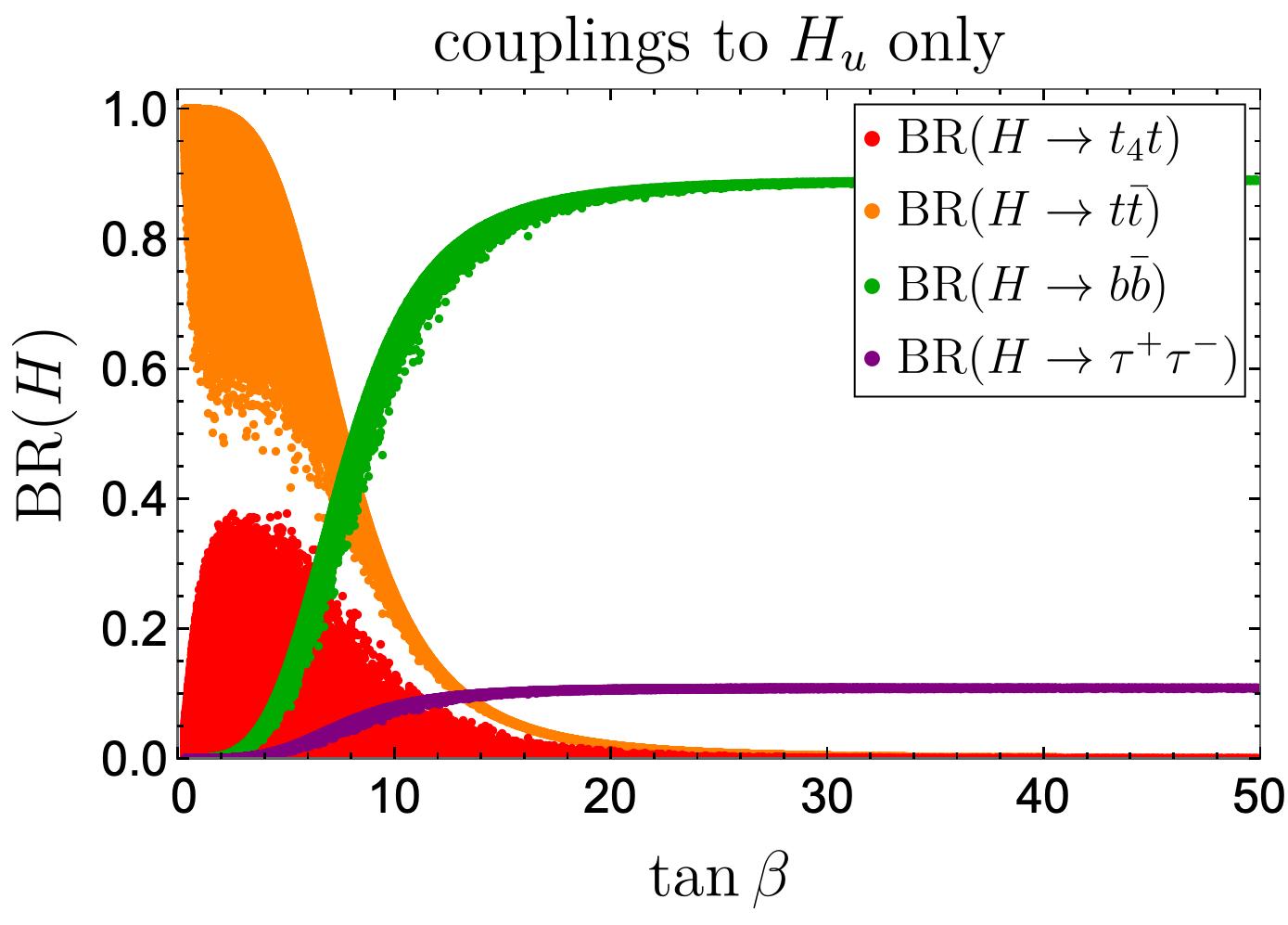}
\includegraphics[width=.49\linewidth]{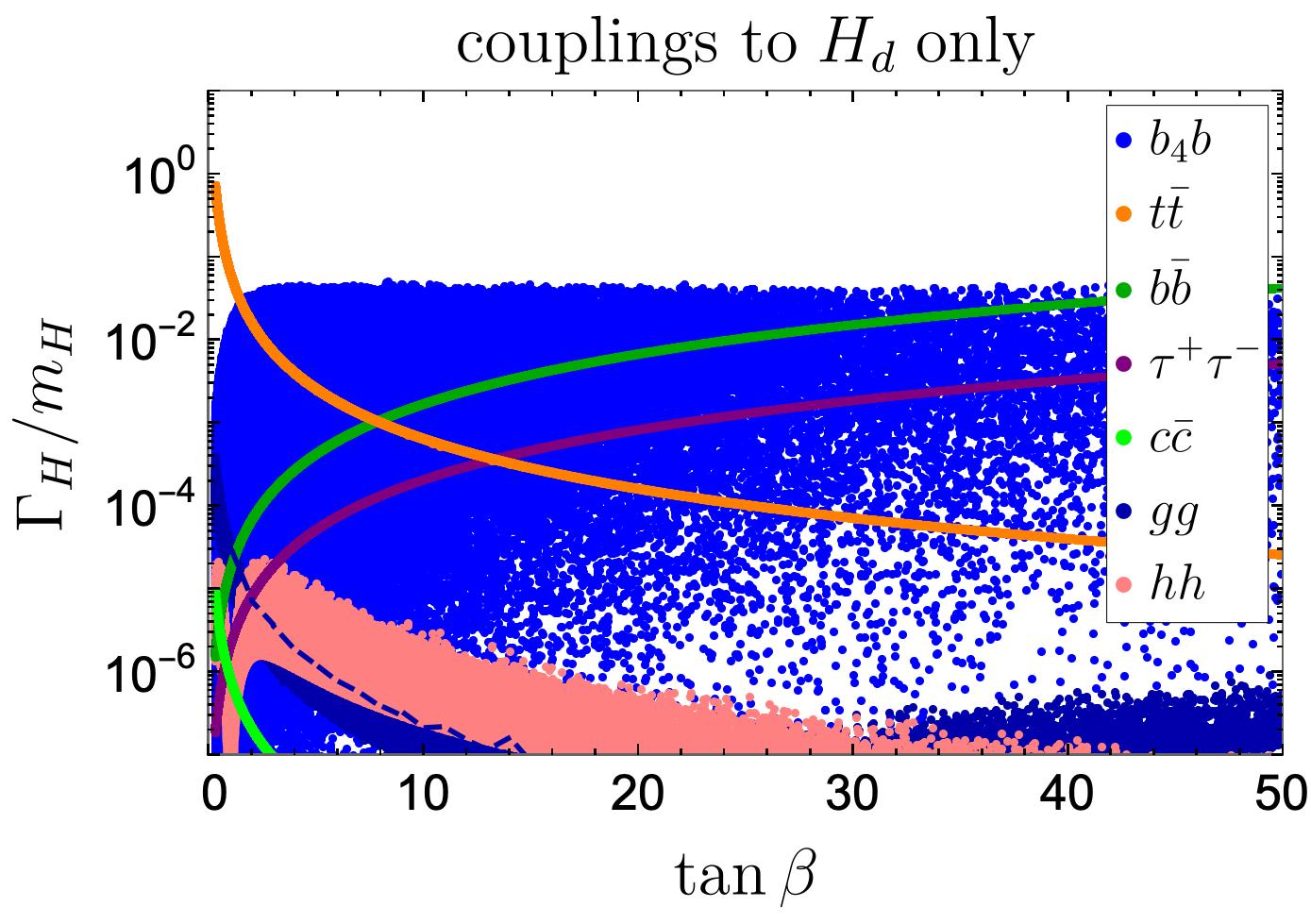}
\includegraphics[width=.49\linewidth]{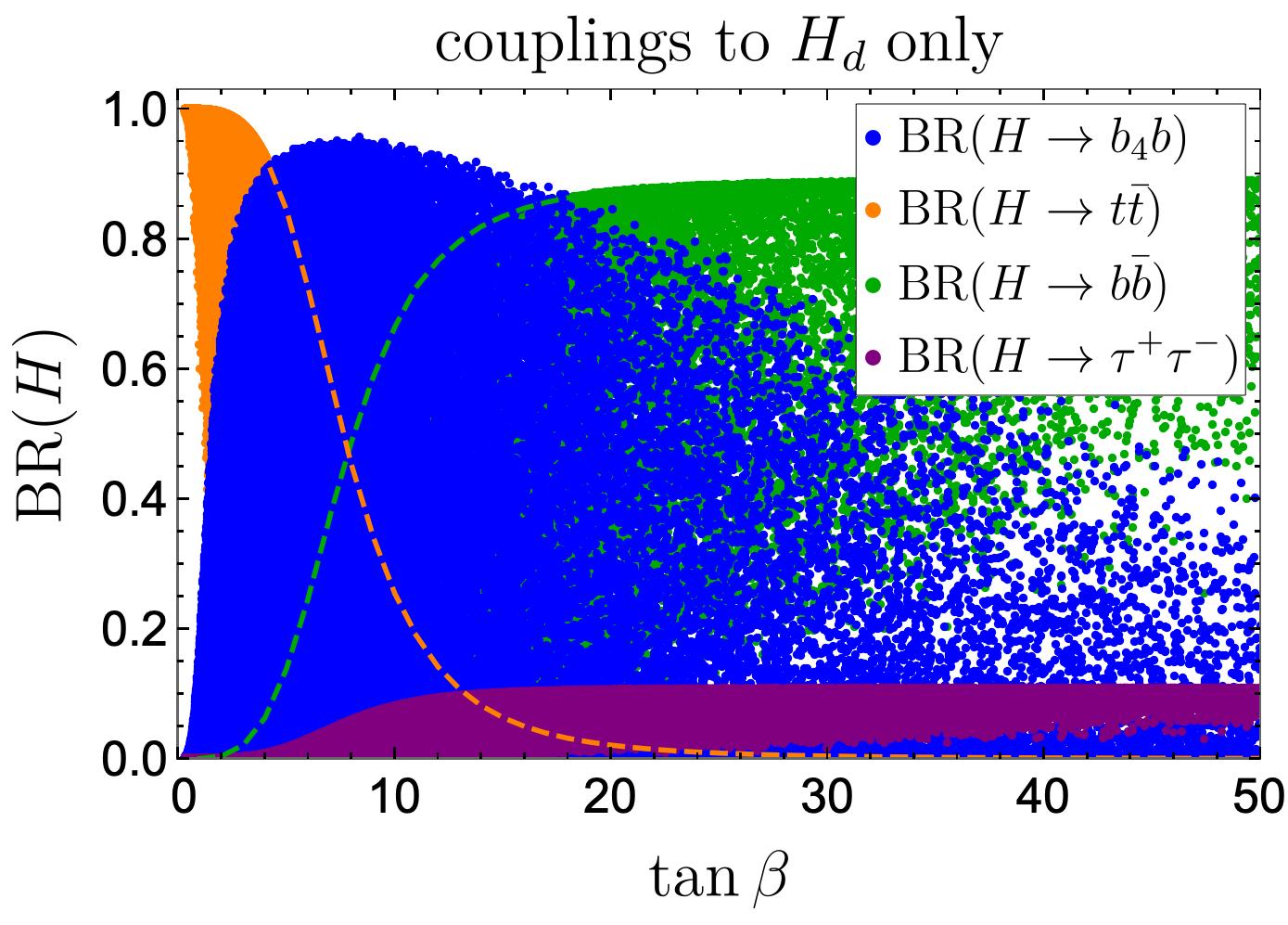}
\includegraphics[width=.49\linewidth]{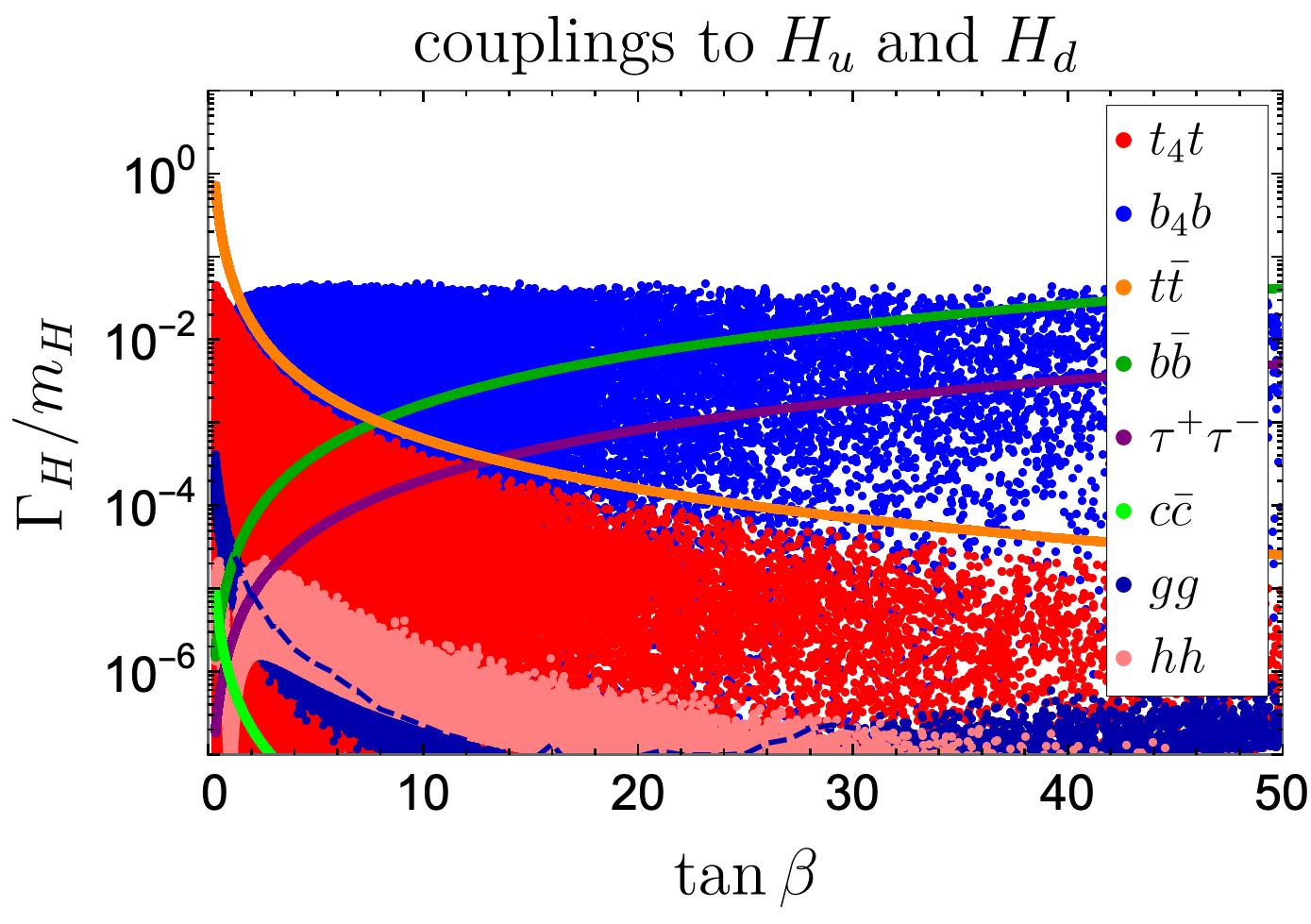}
\includegraphics[width=.49\linewidth]{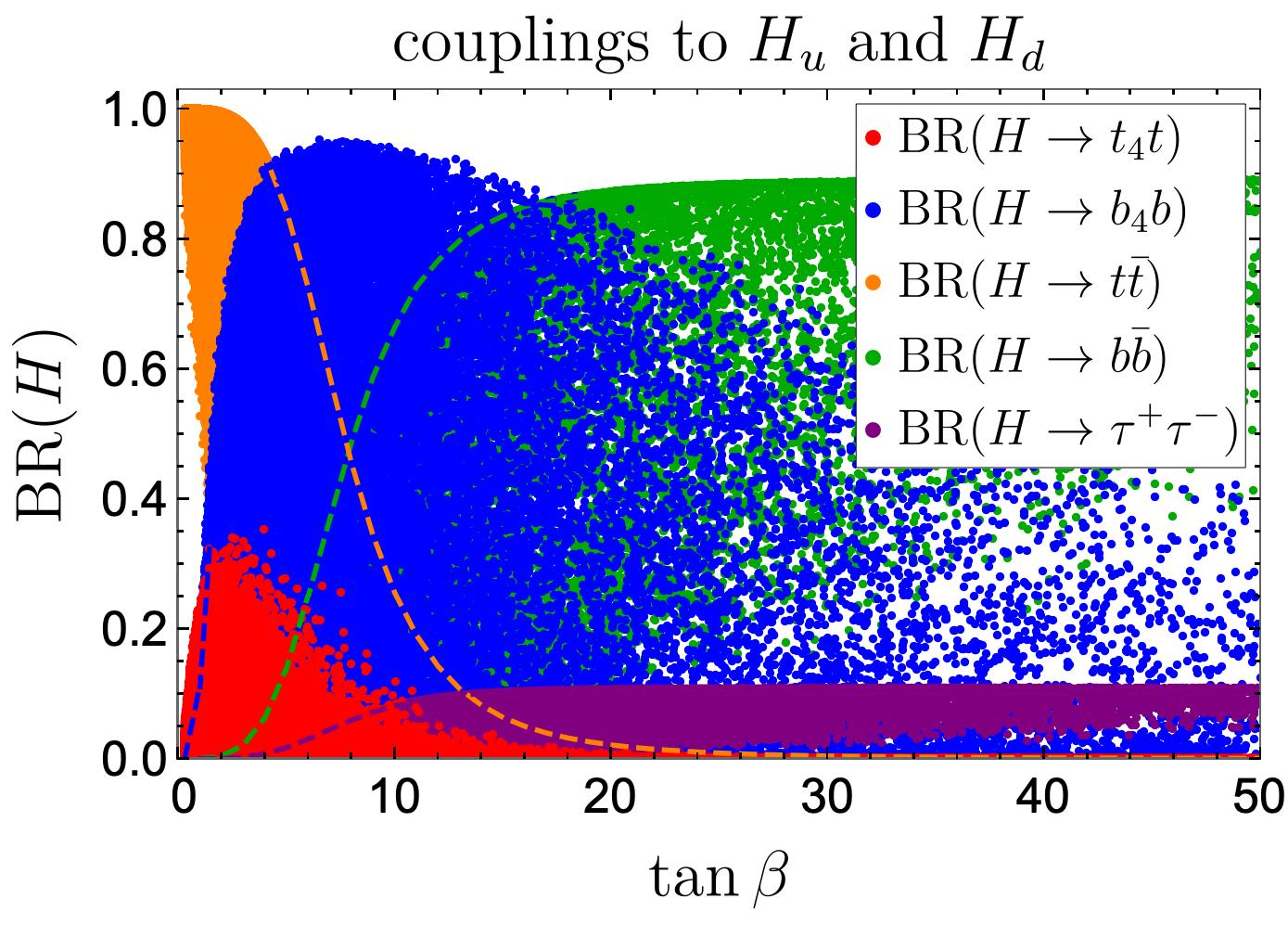}
\caption{Partial widths (left) and branching ratios (right) of a heavy neutral Higgs boson as functions of $\tan \beta$ assuming couplings to $H_u$ only (top), $H_d$ only (middle), and $H_u$ and $H_d$ (bottom).}
\label{fig:Hdecays}
\end{center}
\end{figure}

In figure~\ref{fig:Hdecays} we show the $\tan\beta$ dependence of Higgs  partial decay widths and branching ratios for scenarios with couplings to $H_u$ only 
(all $\lambda$s = 0), $H_d$ only 
(all $\kappa$s = 0)
and to both $H_u$ and $H_d$. The dominant features of these plots can be easily understood from the $\tan\beta$ dependence of the heavy Higgs couplings to $t$ and $t_4$: $\lambda^H_{t t} \propto 1/\tan\beta$ and $\lambda^H_{t_4 t} \propto \cos\beta$, and couplings to $b$ and $b_4$:  $\lambda^H_{b b} \propto \tan\beta$ and $\lambda^H_{b_4 b} \propto \sin\beta$  (see table~2 of Ref.~\cite{Dermisek:2019vkc}). These $\tan\beta$ dependences directly translate into the dependence of the partial widths in the left panels of figure~\ref{fig:Hdecays}.

In the scenario with couplings to $H_u$ only, the $H\to t_4 t$ mode is asymptotically smaller than both 
$H\to t \bar t$ 
(at small $\tan\beta$) and 
$H\to b \bar b$ 
(at large $\tan\beta$) and is relevant only at small-to-medium $\tan\beta$: we find branching ratios larger than 10\% for $\tan\beta \in [0.5,10]$ and they can reach up to $40\%$. On the other hand, in the scenario with couplings to $H_d$, the $H\to b_4 b$ mode is still asymptotically smaller than 
$H\to t \bar t$ 
at small $\tan\beta$ but is not suppressed with respect to 
$H\to b \bar b$ 
at large $\tan\beta$. We find branching ratios larger than 10\% for any $\tan\beta > 0.8$. More importantly, this mode can dominate for $\tan\beta\in [4,18]$ and can reach up to 95\%. The scenario with couplings to both $H_u$ and $H_d$ can be understood in a similar way.\footnote{In the calculation of the branching ratios we account for decays through both $t_4 (b_4)$ and $t_5 (b_5)$ when kinematically open. This effectively reduces the branching ratio into $t_4 (b_4)$.}

Note that the maximum Higgs partial widths and branching ratios into vectorlike quarks depend on the ranges of Yukawa couplings that we scan over that are given in Eqs.~(\ref{rangek}) and (\ref{rangel}). The maximum partial widths scale with the square of the maximum coupling allowed; for example, limiting the upper ranges to 0.5 reduces the maximum widths by a factor of 4. The impact on the branching ratios is less straightforward. Reducing the upper range of the scan to 0.5 implies that the $H \to t_4 t$ branching ratio peaks at 15\%; the $H \to b_4 b$ branching ratio dominates for $\tan\beta \in[5,10]$, peaks at about 85\% but drops to about 20\% for $\tan\beta \sim 50$.

\begin{figure}
\begin{center}
\includegraphics[width=.7\linewidth]{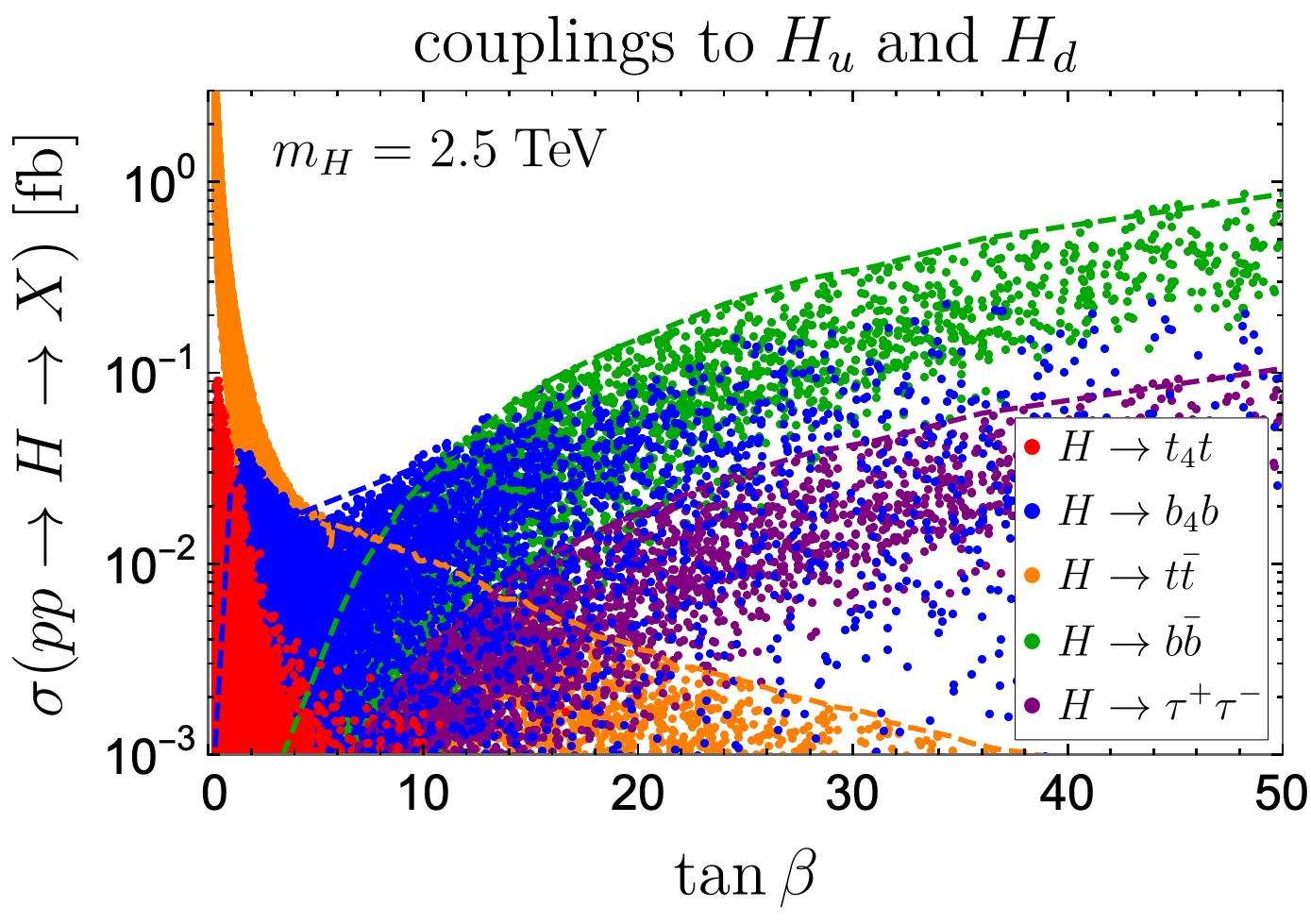}
\caption{Production cross section of a heavy neutral Higgs boson multiplied by branching ratios to individual decay modes as functions of $\tan \beta$ for $m_H = 2.5$ TeV.}
\label{fig:Hproddecay}
\end{center}
\end{figure}

Due to different $\tan\beta$ dependence of  Higgs production cross section and branching ratios, it is interesting to show the total rate into individual final states. In figure~\ref{fig:Hproddecay} we show the various production rates for $m_H = 2.5 \; \text{TeV}$ as functions of $\tan\beta$. We see that the $t_4 t$ mode is the largest at very small $\tan\beta$ and that, although the $H\to b_4 b$ mode can dominate at medium $\tan\beta$, the $\sigma(pp\to H \to b_4b)$ can still be larger at both  small and very large $\tan\beta$. Rates of the order of 0.1 fb are attainable for $H\to t_4 t$ at very small $\tan\beta$ and for $H\to b_4 b$ at medium-to-large $\tan\beta$.

\begin{figure}
\begin{center}
\includegraphics[width=.32\linewidth]{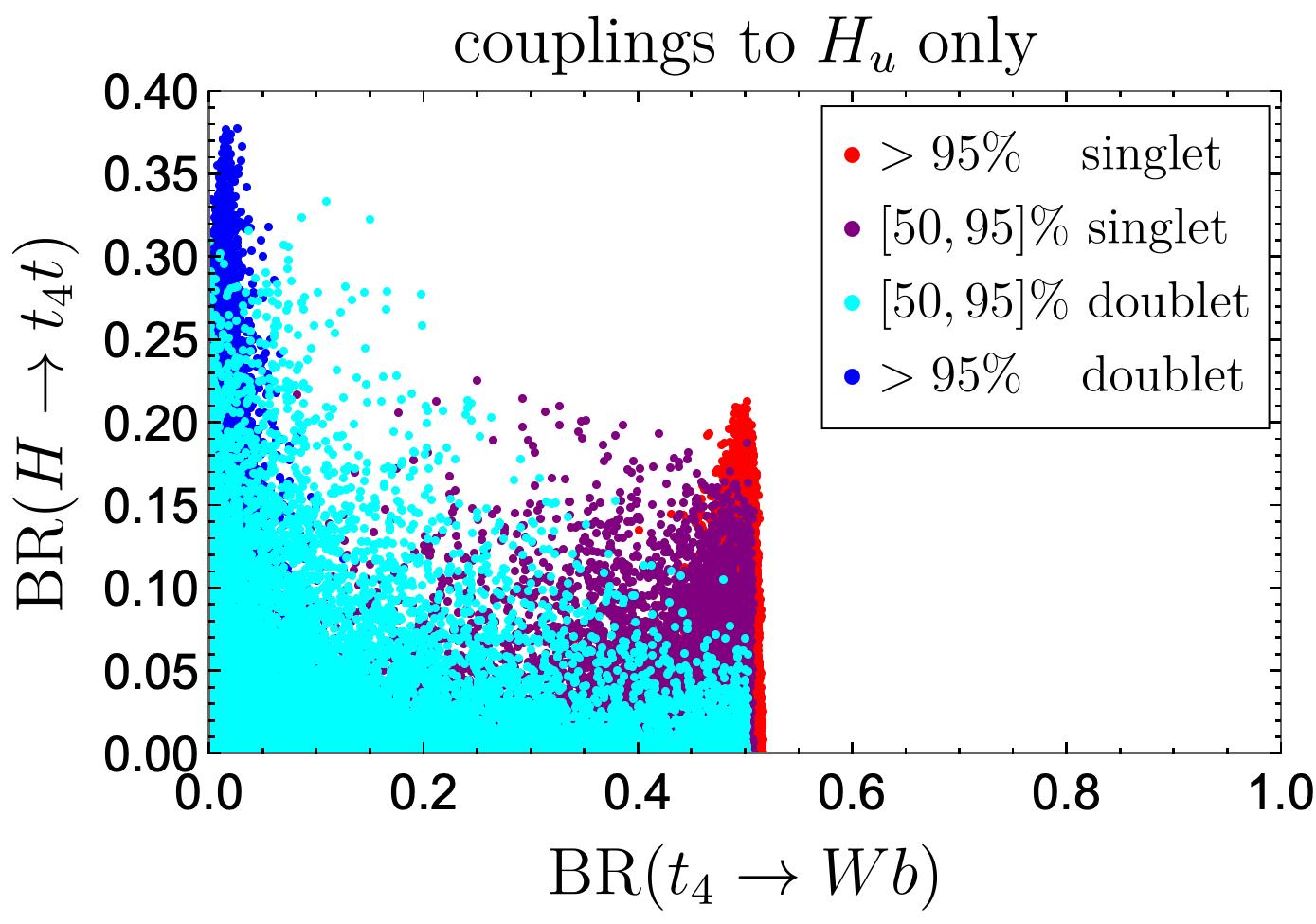}
\includegraphics[width=.32\linewidth]{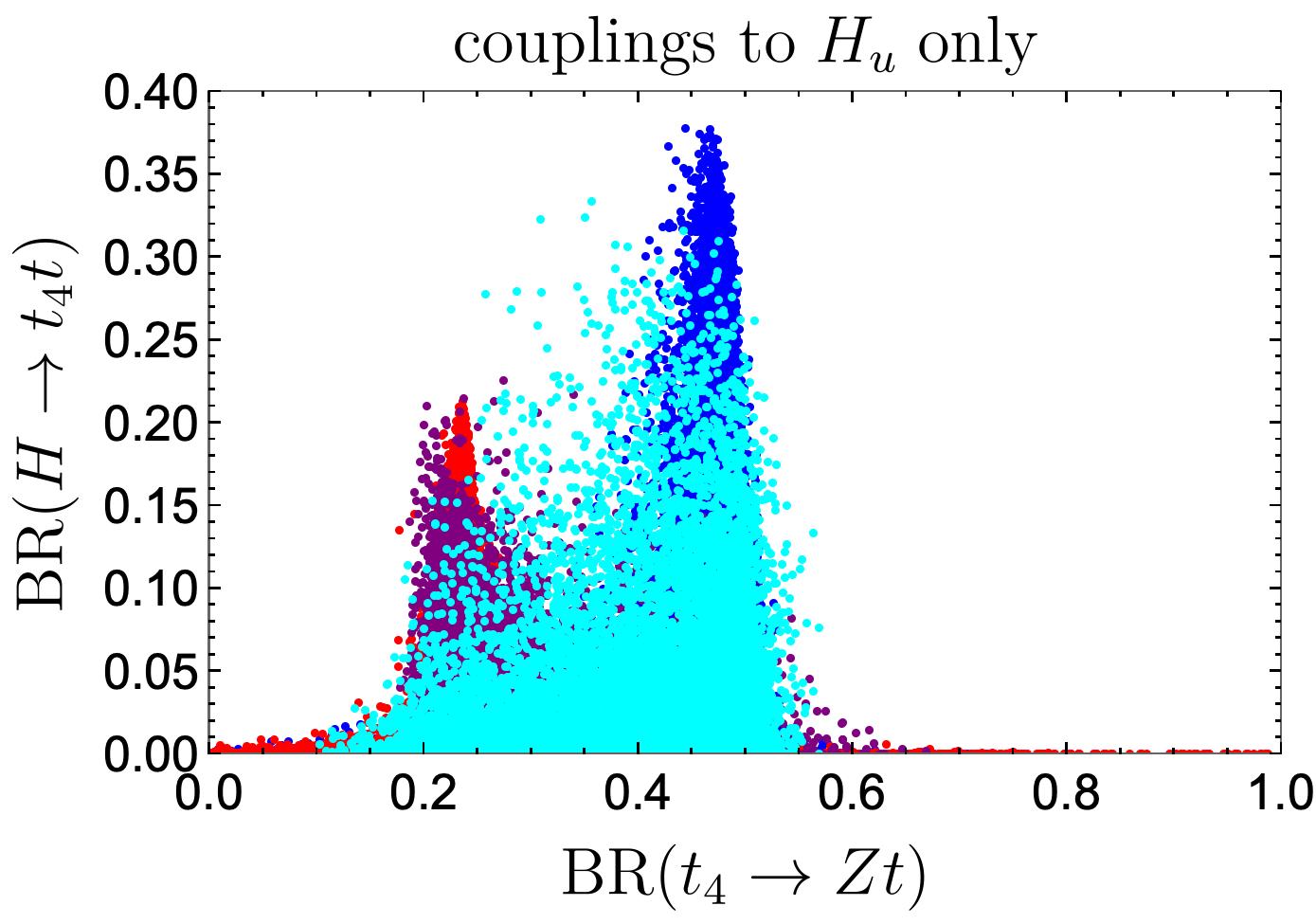}
\includegraphics[width=.32\linewidth]{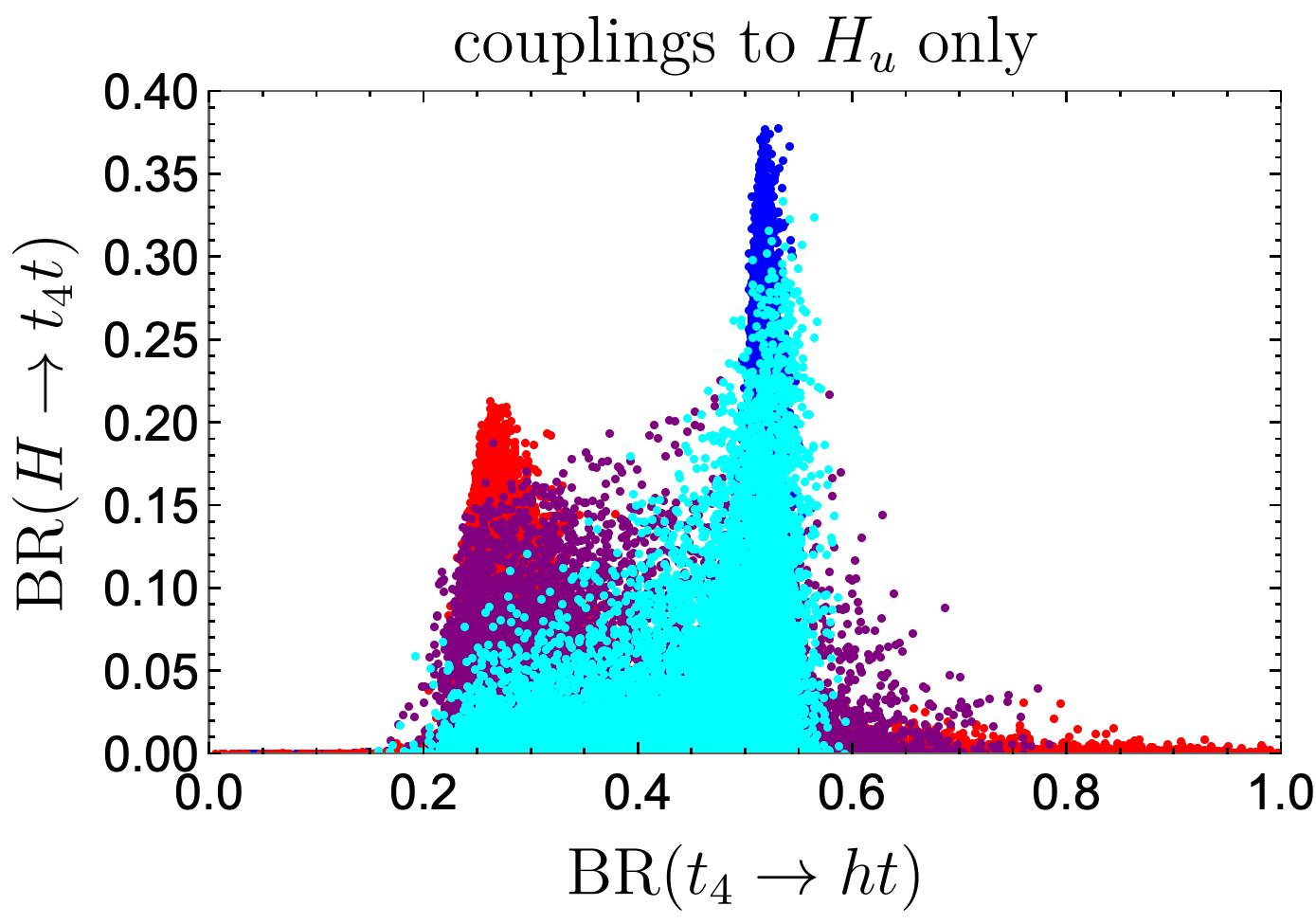}
\includegraphics[width=.32\linewidth]{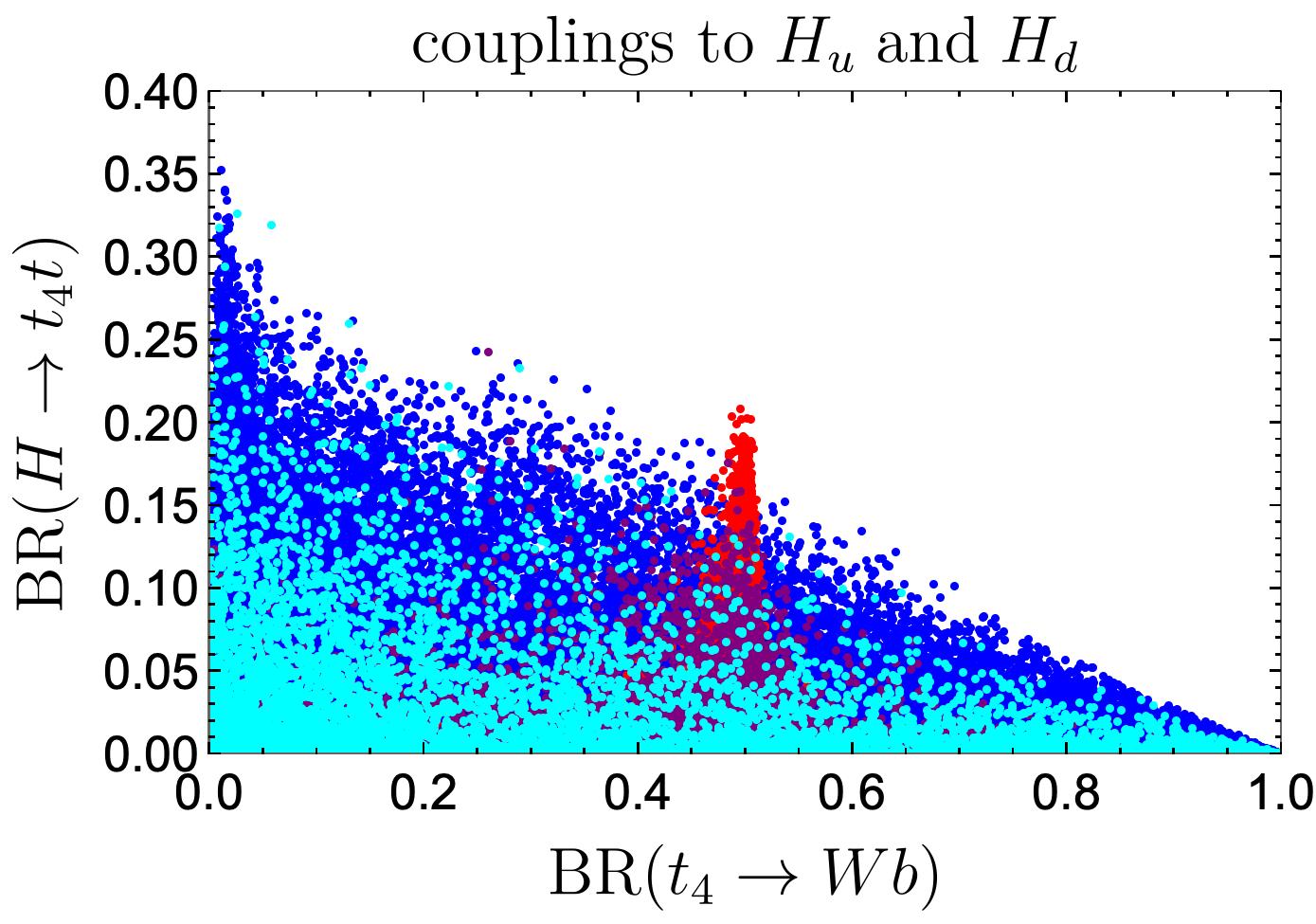}
\includegraphics[width=.32\linewidth]{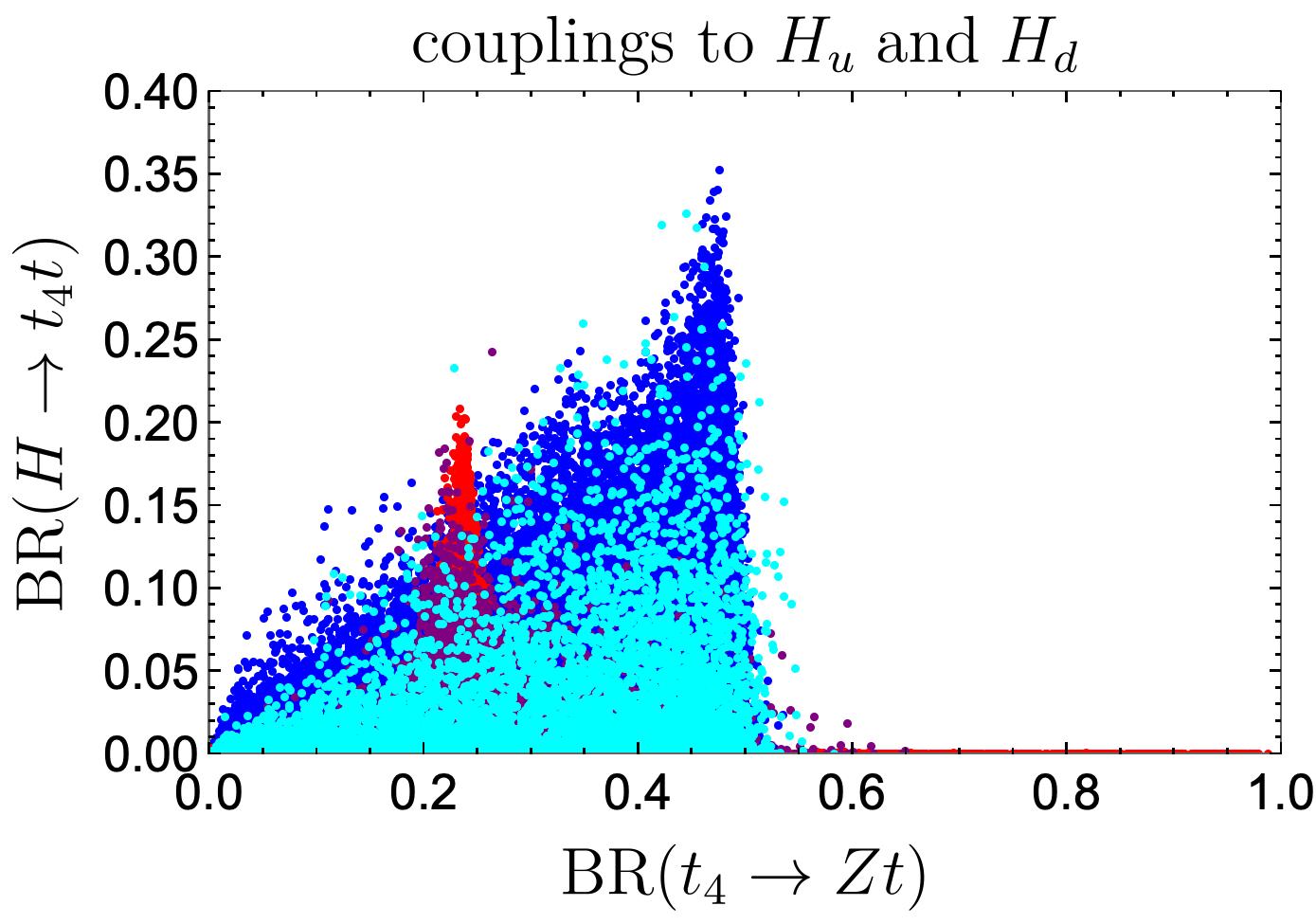}
\includegraphics[width=.32\linewidth]{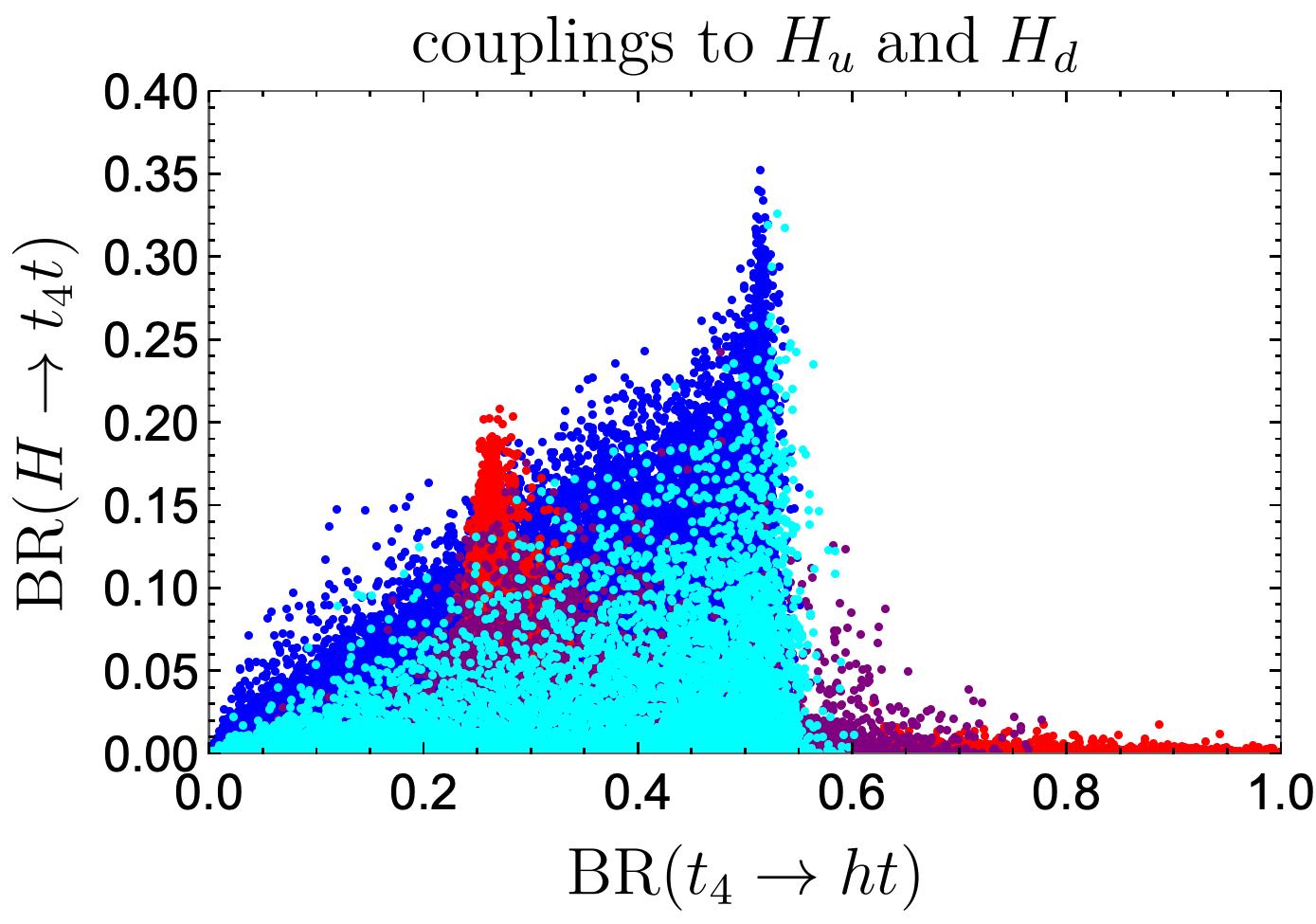}
\caption{Correlations between the branching ratio of $H \to t_4t$ and branching ratios of $t_4$ assuming couplings to $H_u$ only (top) and couplings to $H_u$ and $H_d$ (bottom). Different colors correspond to different singlet/doublet fractions of  $t_4$.}
\label{fig:BRH_vs_BRt4}
\end{center}
\end{figure}

\begin{figure}
\begin{center}
\includegraphics[width=.32\linewidth]{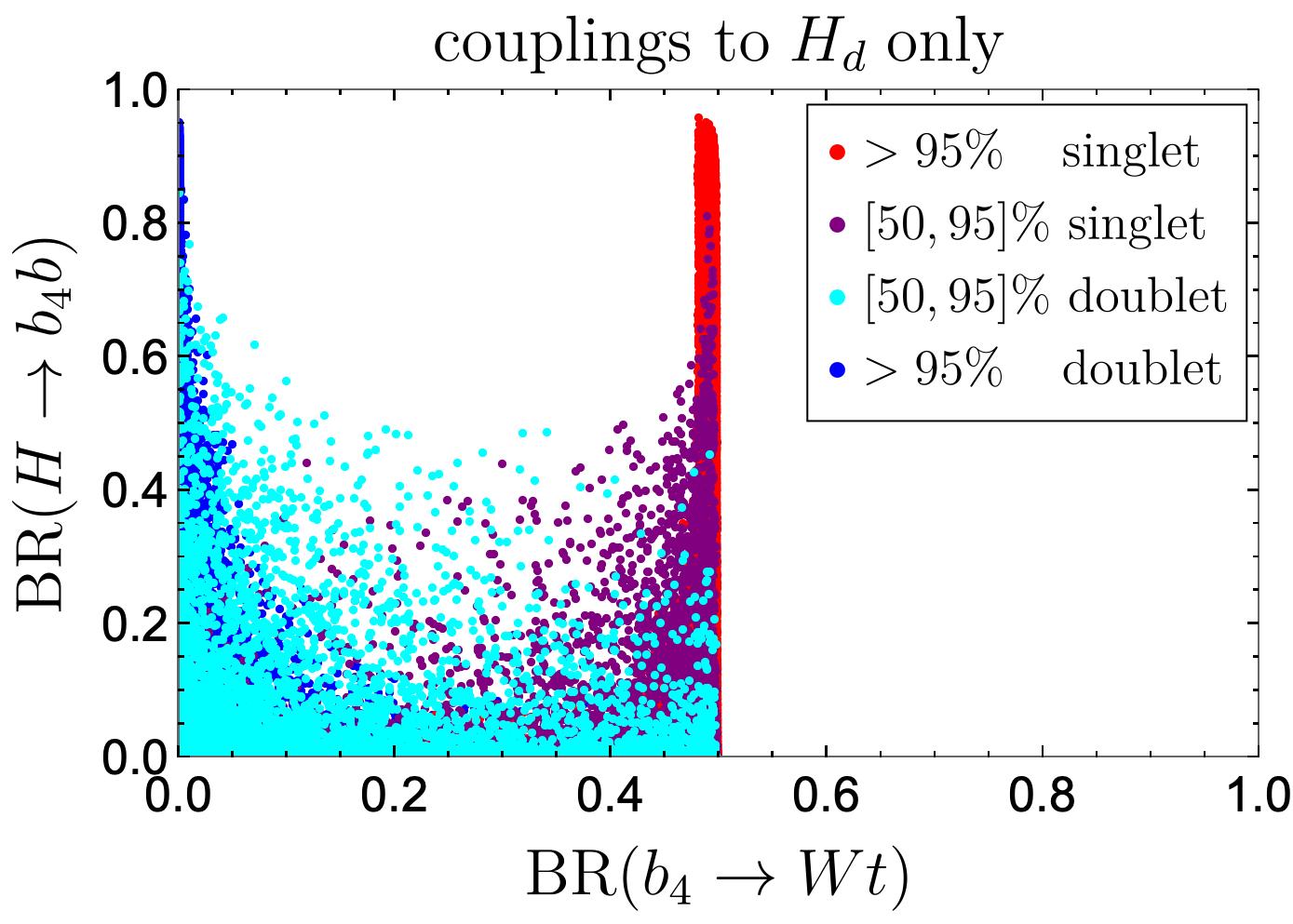}
\includegraphics[width=.32\linewidth]{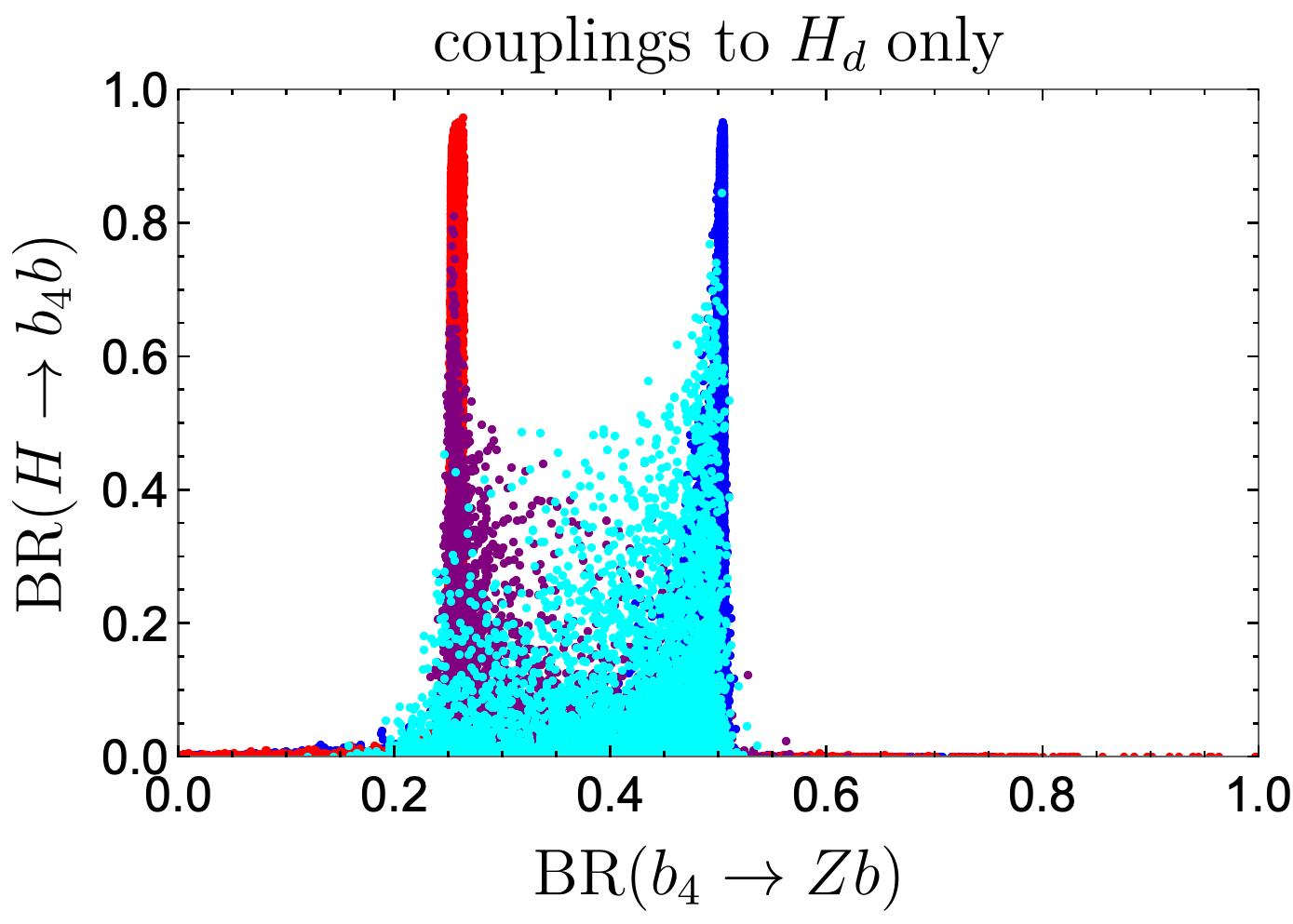}
\includegraphics[width=.32\linewidth]{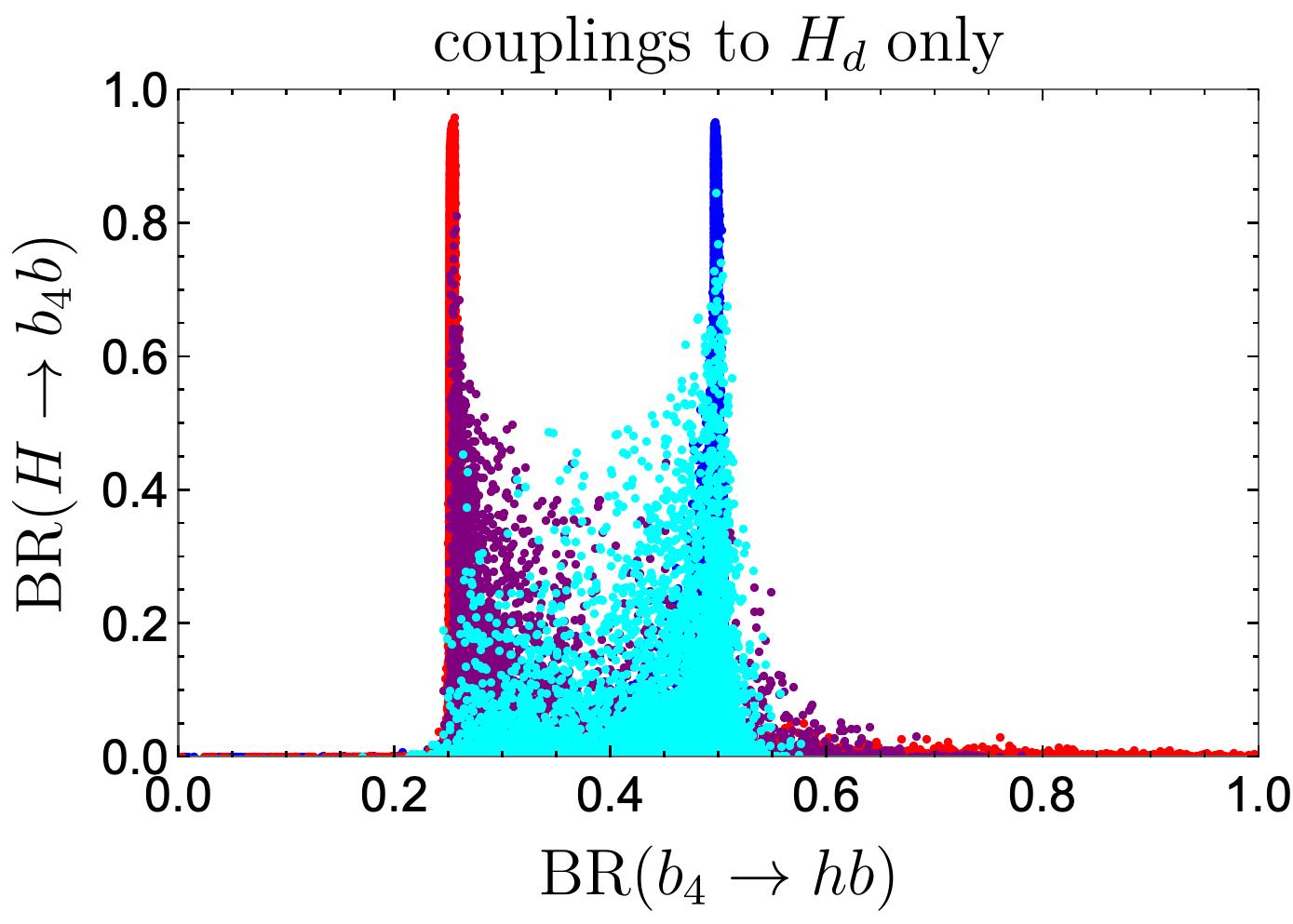}
\includegraphics[width=.32\linewidth]{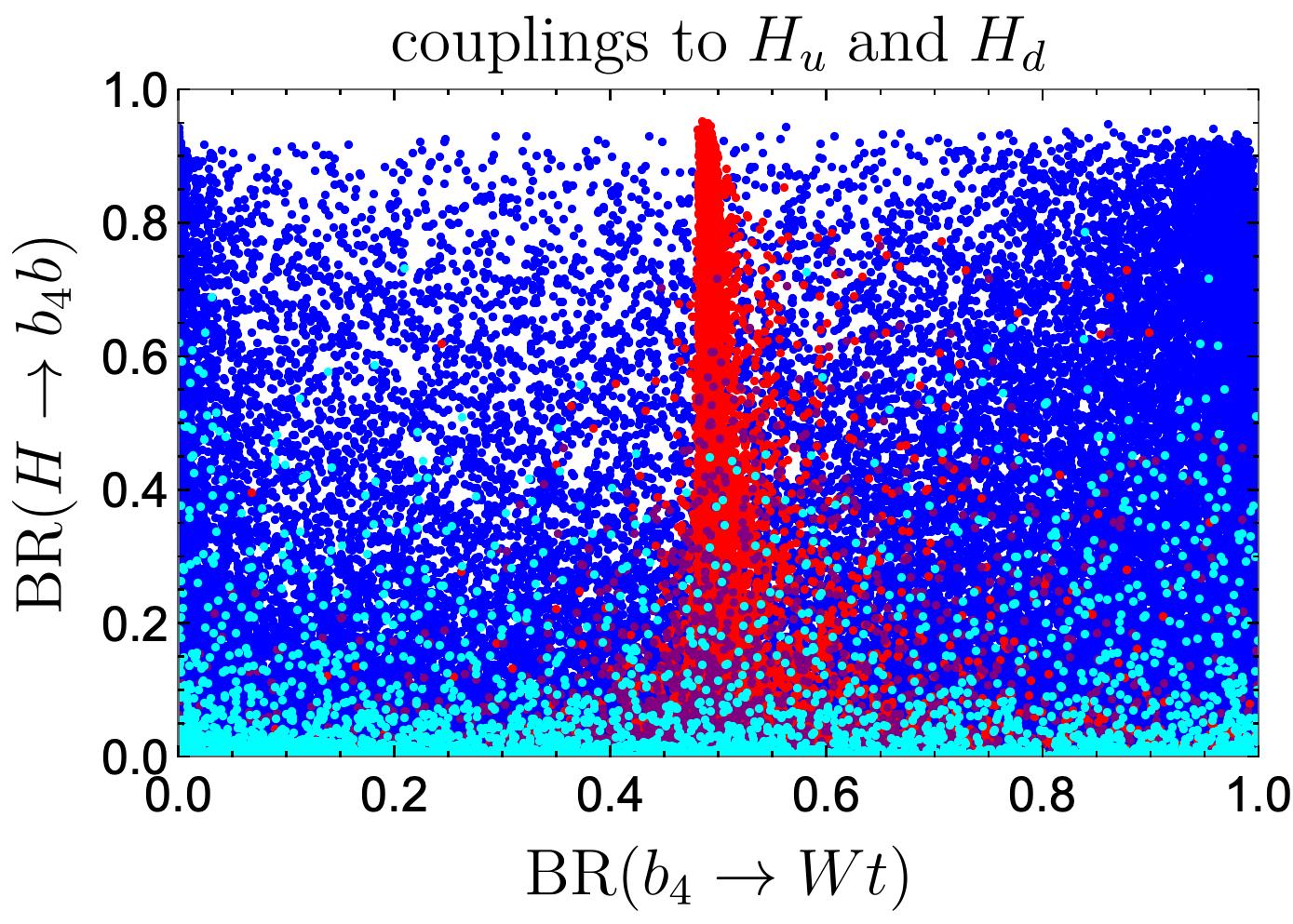}
\includegraphics[width=.32\linewidth]{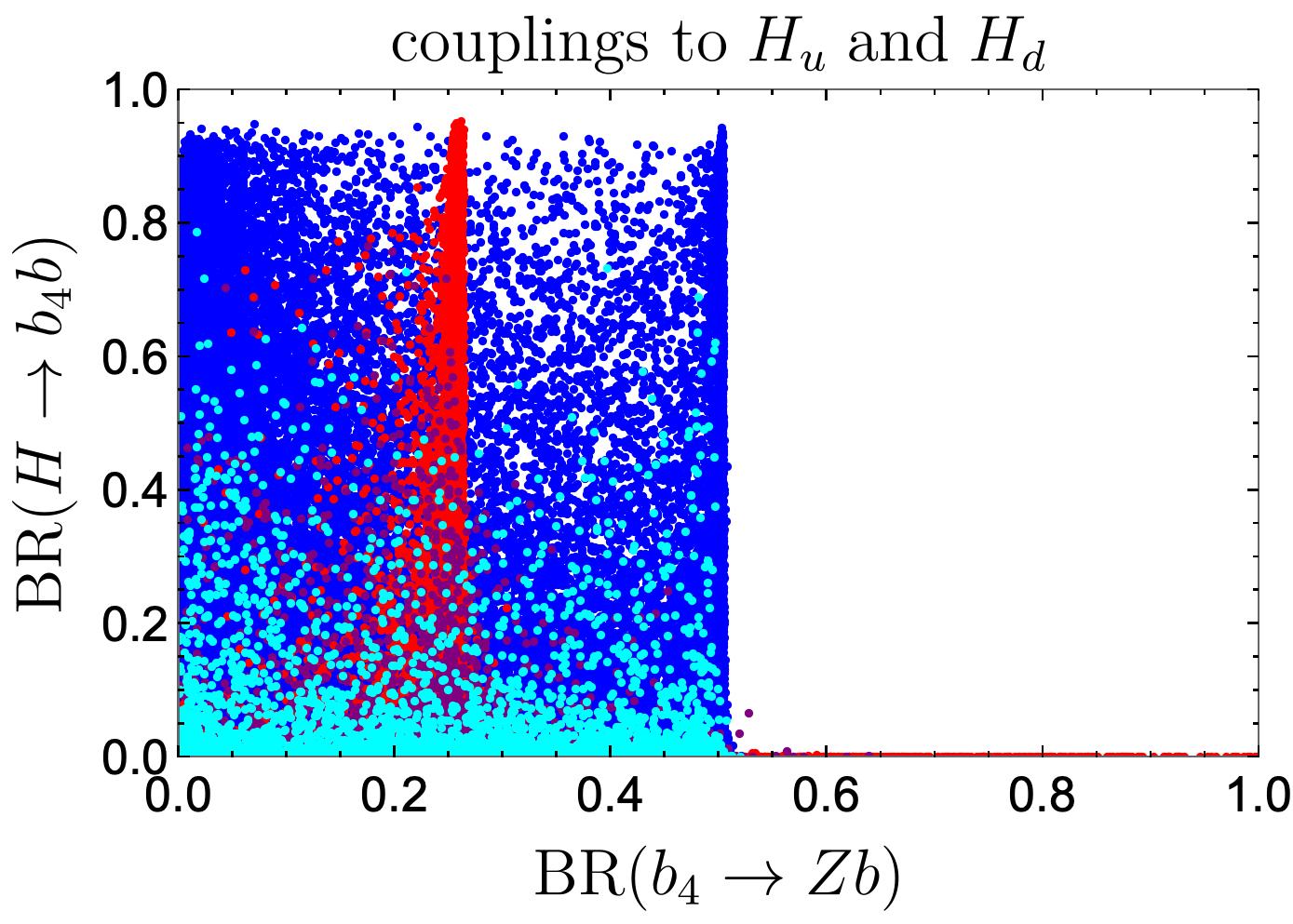}
\includegraphics[width=.32\linewidth]{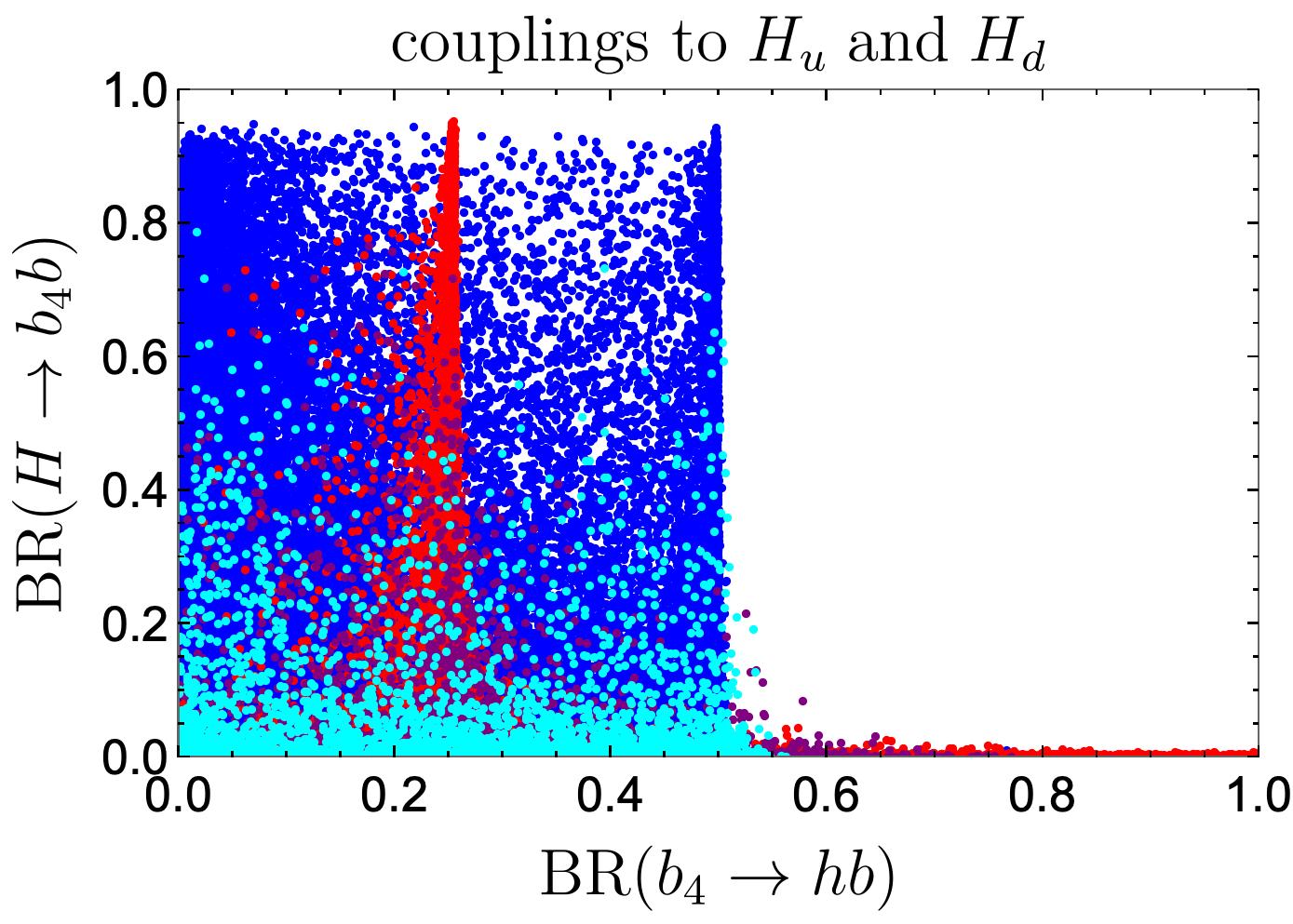}
\caption{Correlations between the branching ratio of $H \to b_4b$ and branching ratios of $b_4$ assuming couplings to $H_d$ only (top) and couplings to $H_u$ and $H_d$ (bottom). Different colors correspond to different singlet/doublet fractions of  $b_4$.}
\label{fig:BRH_vs_BRb4}
\end{center}
\end{figure}

The lightest new quarks from heavy Higgs decays  further decay into SM particles. The correlations between  the branching ratio of $H \to t_4t$ and individual branching ratios of $t_4$ are shown in figure~\ref{fig:BRH_vs_BRt4} and similar correlations for $b_4$ are shown in figure~\ref{fig:BRH_vs_BRb4}. Main features of these plots can be understood from approximate formulas and the discussion in ref.~\cite{Dermisek:2019vkc}.  We see that
the  decay modes of $t_4$ and $b_4$ into $W$, $Z$ and $h$, cluster around the pattern expected from the Goldstone boson equivalence limit corresponding to sending all vectorlike quark masses to infinity. For singlet-like new quarks (red)\footnote{Singlet and doublet fractions of new quarks are defined in ref.~\cite{Dermisek:2019vkc}.}
this leads to 2:1:1 branching ratios into $W$, $Z$ and $h$. For doublet-like new quarks (blue) this leads to a one parameter family of branching ratios characterized by an arbitrary branching ratio to $W$ and equal branching ratios to $Z$ and $h$. For example, for a doublet-like $t_4$, the $W t_4 t$ coupling is controlled by $\lambda_Q$ while the corresponding couplings to $Z$ and both Higgs bosons are controlled by $\kappa_Q$. This   results in a difference between the plots on the top (no couplings to $H_d$ allowed) and bottom (all couplings allowed) in figure~\ref{fig:BRH_vs_BRt4} and similarly for the $b_4$ in figure~\ref{fig:BRH_vs_BRb4}. The main distinction between the corresponding plots in figs.~\ref{fig:BRH_vs_BRt4} and~\ref{fig:BRH_vs_BRb4} originates from different $\tan \beta$ dependence of relevant couplings. Note especially that while the branching ratios for $t_4 \to Wb$ and $b_4 \to W t$ extend to 100\% in the scenario with couplings to both $H_u$ and $H_d$, the former anticorrelates with the $H\to t_4 t$ branching ratio as can be seen in the lower-left panel of figure~\ref{fig:BRH_vs_BRt4}.

The mixed  scenarios (cyan and purple) interpolate between mostly singlet and mostly doublet cases. Note that these scenarios require careful choices of model parameters especially for $b_4$ at medium to large $\tan\beta$, see eq.~(\ref{eq:mmd}), where $H\to b_4 b$ is sizable. This is the reason for an empty area in  between the mostly singlet and mostly doublet cases in the top plots of figure~\ref{fig:BRH_vs_BRb4}. It is expected that with large statistics the whole area would be populated.

Finally, as discussed in detail in ref.~\cite{Dermisek:2019vkc}, with  the general structure of Yukawa matrices that we allow,  essentially arbitrary branching ratios  of $t_4$ and $b_4$ can be achieved. However, going away from the Goldstone boson equivalence limit correlates with diminishing $H\to t_4t$ and $H\to b_4 b$ because it requires very small $\kappa_{Q,T}$ and $\lambda_{Q,B}$  couplings that are directly related to $H t_4 t$ and $H b_4 b$ couplings.

\begin{figure}
\begin{center}
\includegraphics[width=.32\linewidth]{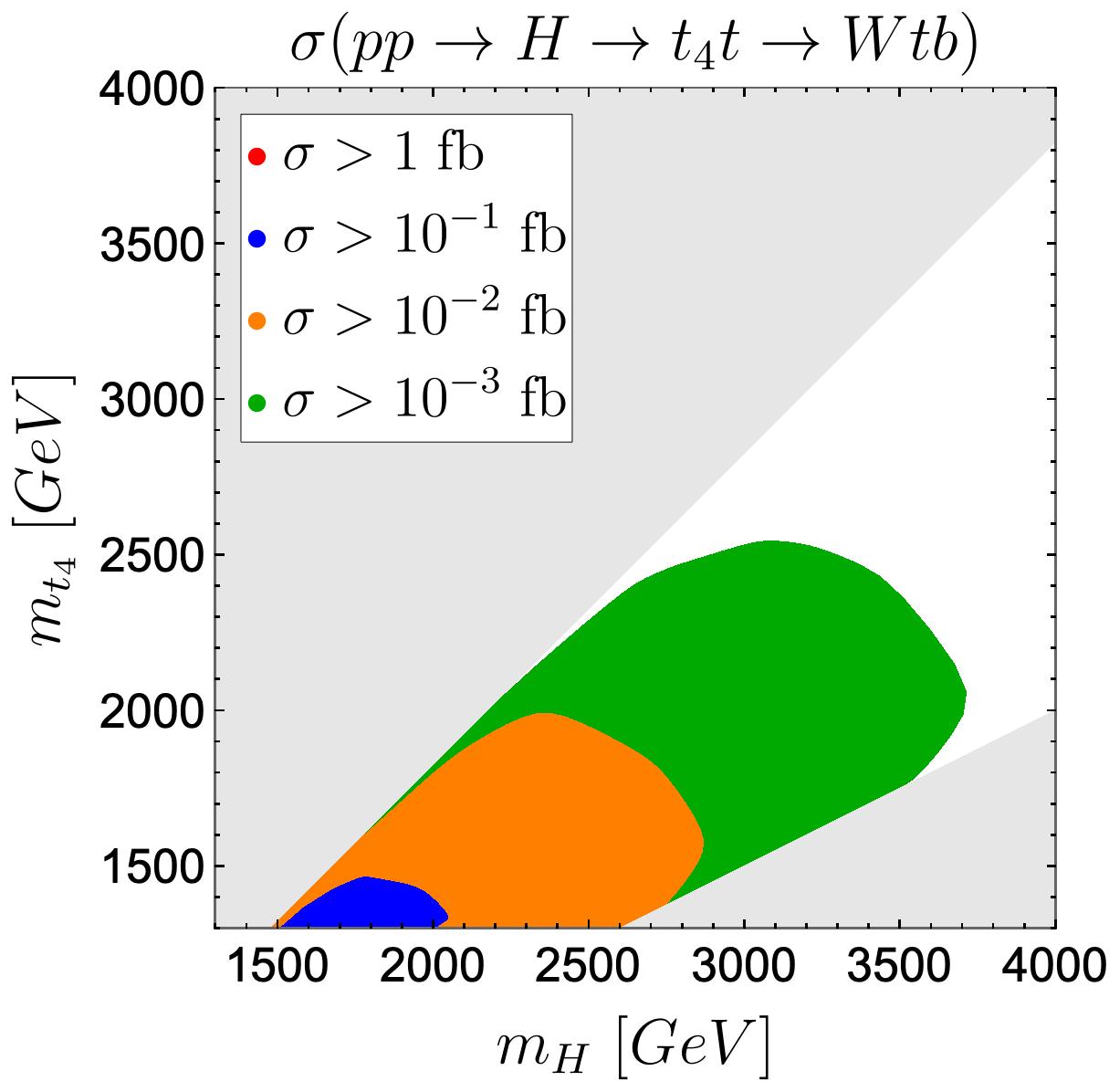}
\includegraphics[width=.32\linewidth]{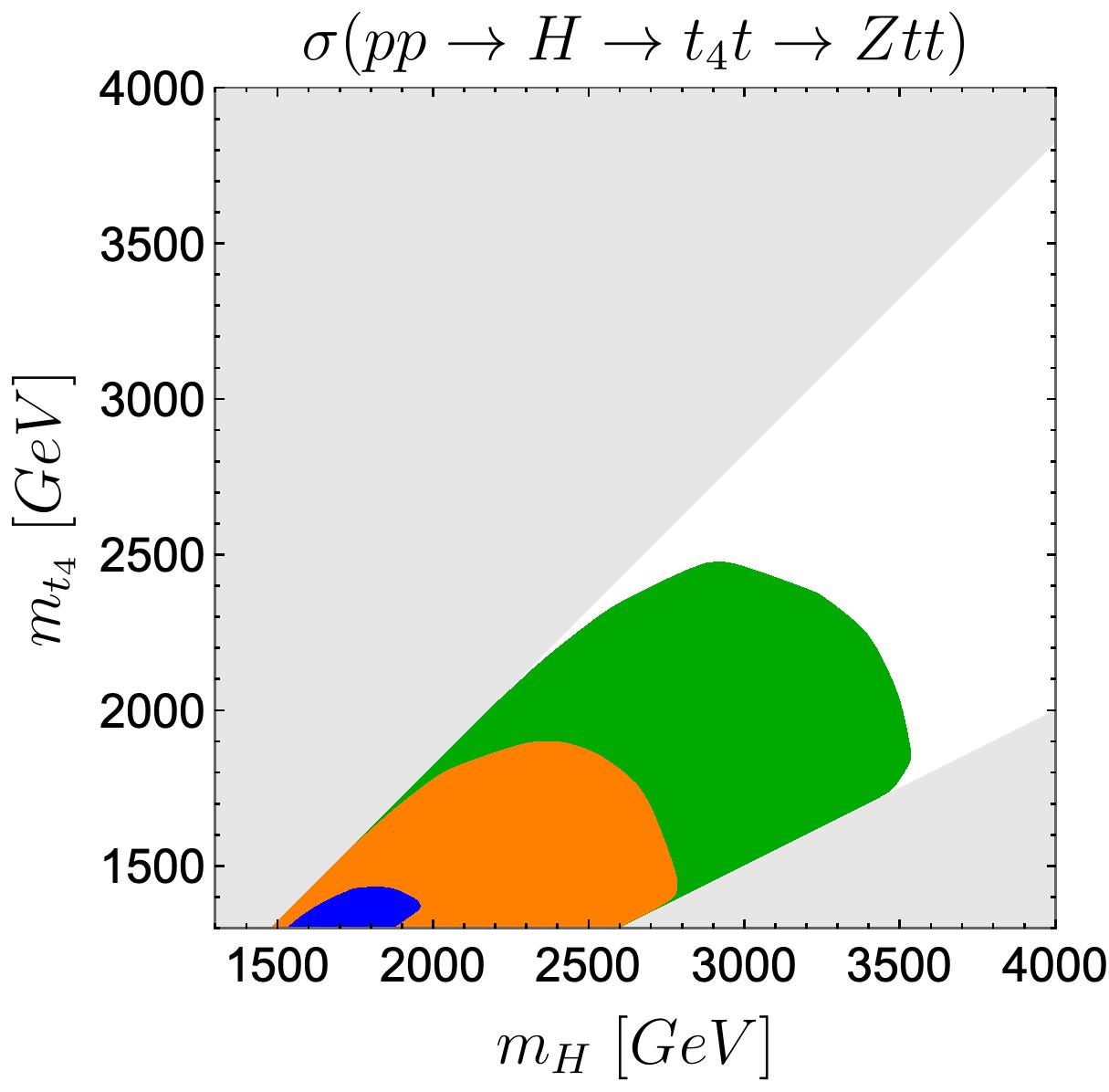}
\includegraphics[width=.32\linewidth]{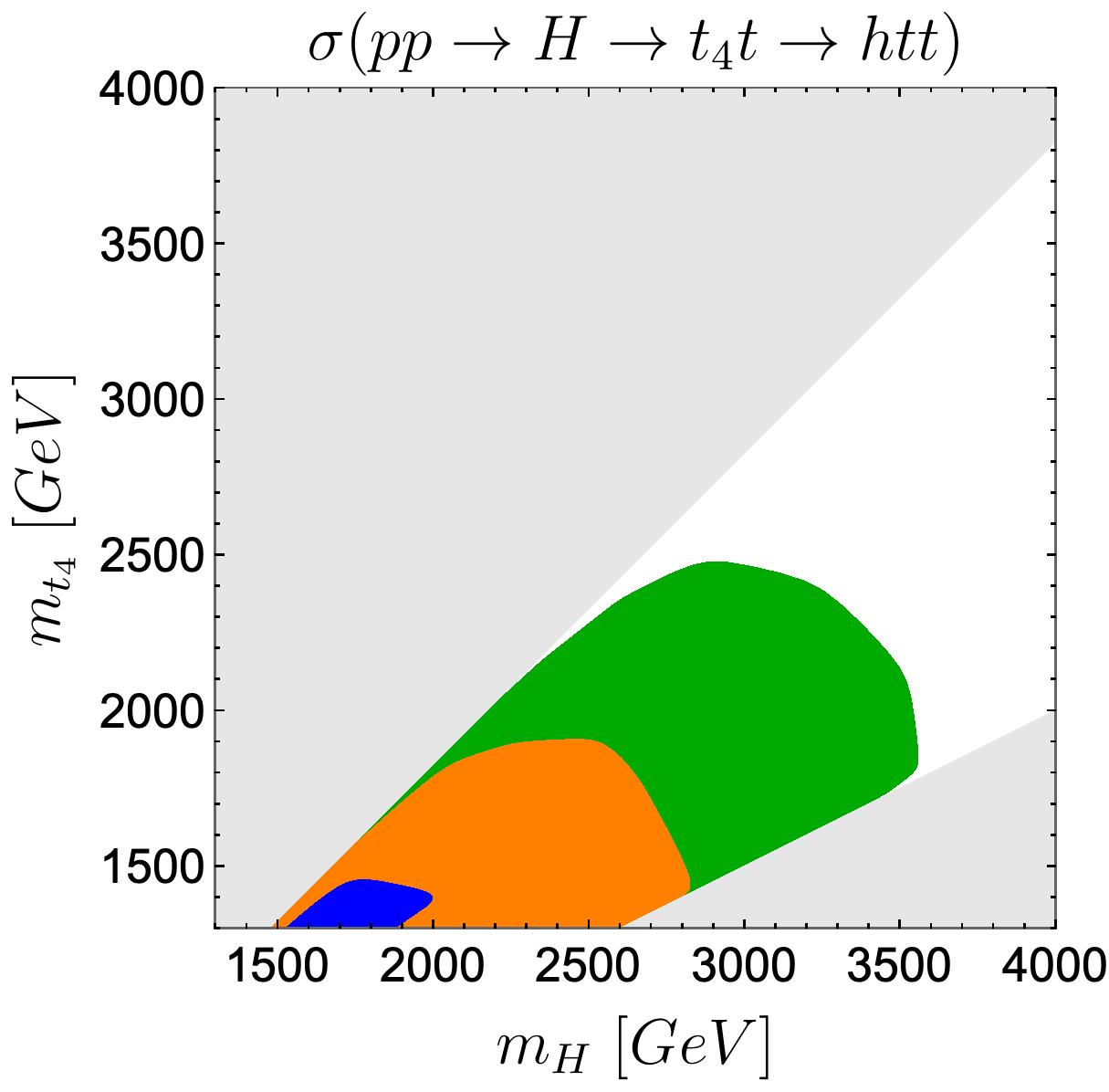}
\includegraphics[width=.32\linewidth]{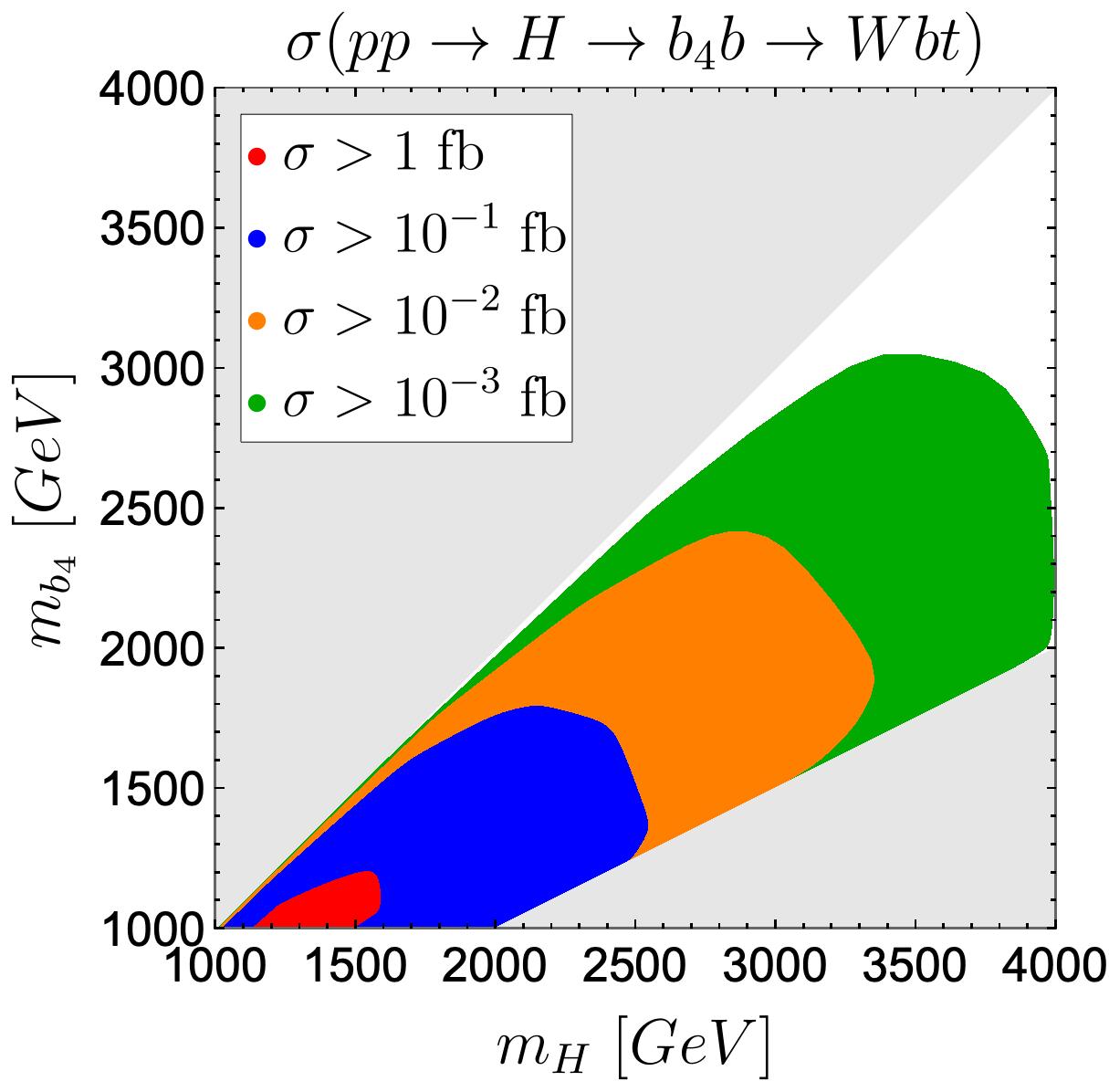}
\includegraphics[width=.32\linewidth]{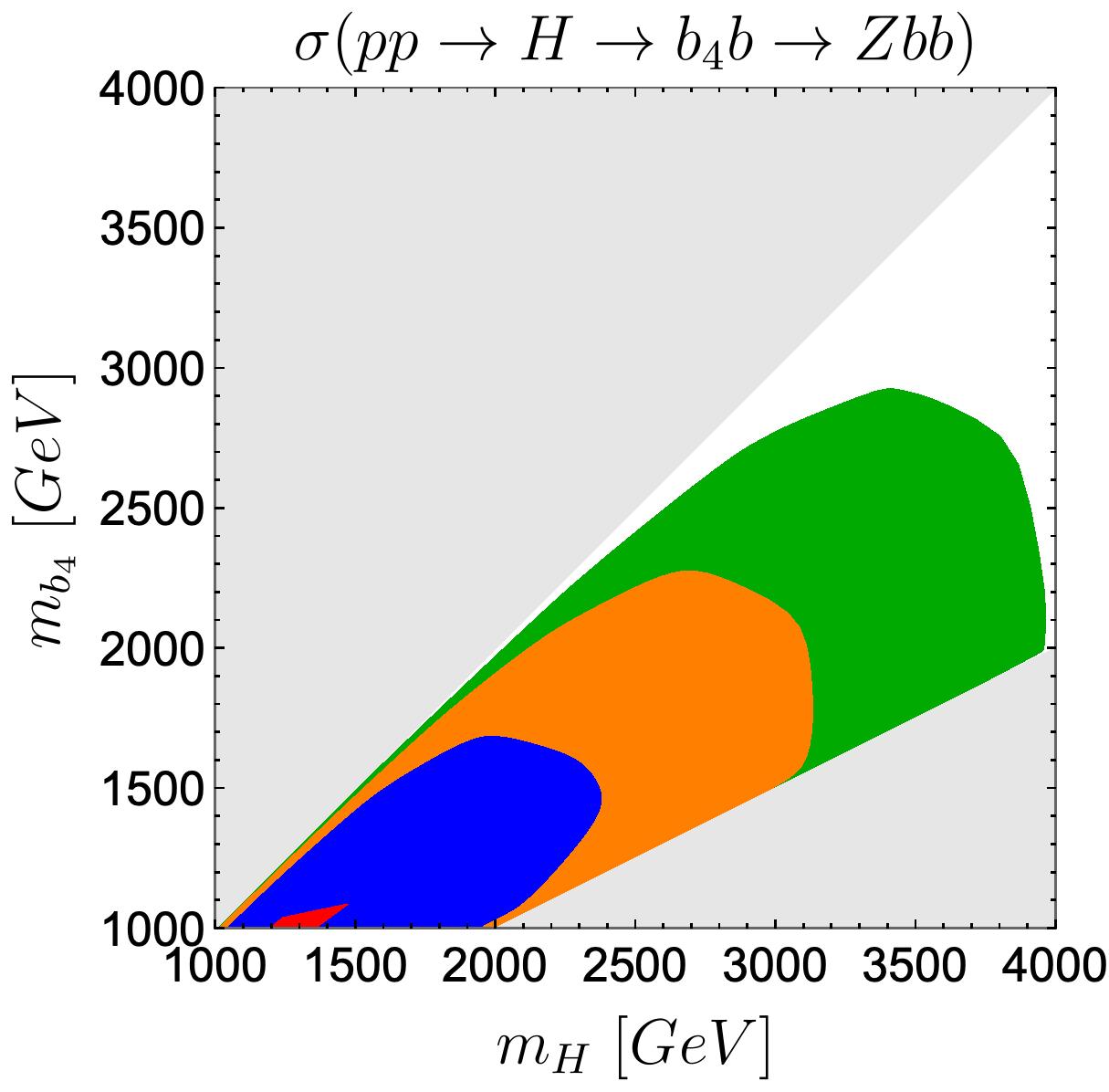}
\includegraphics[width=.32\linewidth]{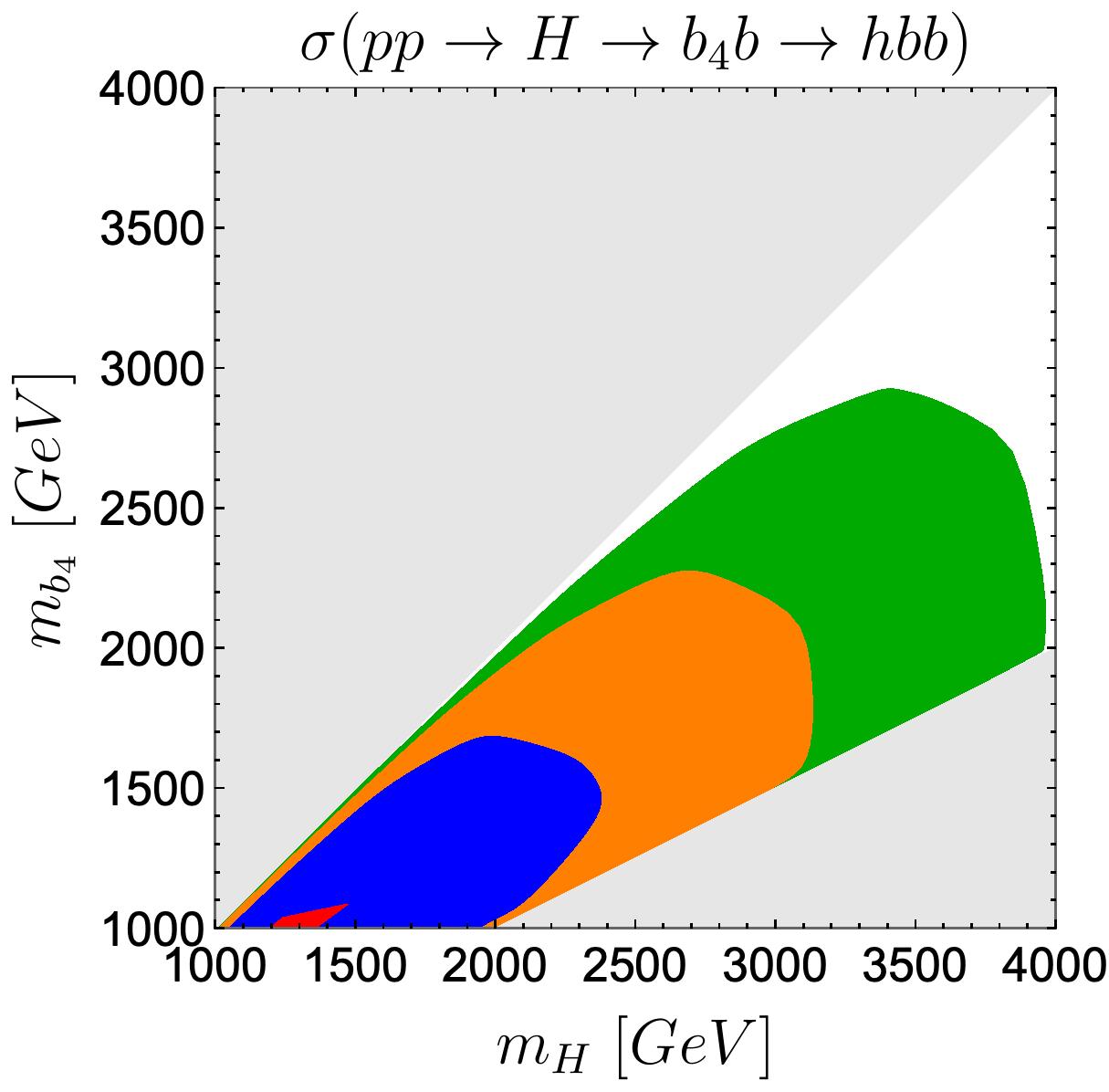}
\caption{Maximum production rates of individual final states in cascade decays of a heavy Higgs boson as functions of $m_H$ and $m_{t_4}$ (top panels) or $m_{b_4}$ (bottom panels) in the scenario with couplings to both $H_u$ and $H_d$. In gray shaded regions $H\to t_4t, \, b_4b$ modes are not kinematically allowed or $H\to t_4 \bar t_4, \, b_4 \bar b_4$ modes are open.}
\label{fig:maxXS}
\end{center}
\end{figure}

The maximum production rates of individual final states in cascade decays of a heavy Higgs boson as  functions of $m_H$ and $m_{t_4}$ or $m_{b_4}$ are presented in figure~\ref{fig:maxXS}. We see that, for Higgs cascade decays through a $t_4$,  rates of 0.1 fb extend up to $m_H \lesssim 2 \; \text{TeV}$ and $m_{t_4} \lesssim 1.4 \; \text{TeV}$; rates above 1 ab can be achieved for $m_H \lesssim 3.5\; \text{TeV}$ or $m_{t_4} \lesssim 2.5 \; \text{TeV}$. For Higgs cascade decays through a $b_4$, rates of individual final states larger than 0.1 fb extend up to $m_H \lesssim  2.5\; \text{TeV}$ and $m_{b_4} \lesssim 1.8 \; \text{TeV}$ and can be even larger than 1 fb for $m_H \lesssim  1.6\; \text{TeV}$ and $m_{b_4} \lesssim 1.2 \; \text{TeV}$; rates above 1 ab can be achieved for $m_H \lesssim 4\; \text{TeV}$ or $m_{b_4} \lesssim 3 \; \text{TeV}$.

\section{Search strategies and  comparison with single productions of $t_4$ and $b_4$}
\label{sec:Searches}
The signatures of cascade decays of a heavy Higgs boson through vectorlike quarks are almost identical to the production of any new resonance (e.g. $Z^\prime$) decaying to $t_4 t$ or $b_4 b$.\footnote{Similar signatures also appeared in the context of composite Higgs models \cite{Greco:2014aza,Backovic:2016ogz} and extra dimensions \cite{Bini:2011zb}.} In addition, the final states we consider are very similar to the single production of both top-like and bottom-like vectorlike quarks, therefore all searches for a singly produced vectorlike quark can be reinterpreted as bounds on Higgs cascade decays. Note however that the topology of cascade decays provides more handles. For example, in the $t_4$ case, there is a top quark in the decay chain, and, in all cases, there is an additional resonance at the heavy Higgs mass. Thus, dedicated searches have a potential to considerably improve the limits found in standard single production studies.

\begin{figure}
\begin{center}
\includegraphics[width=.49\linewidth]{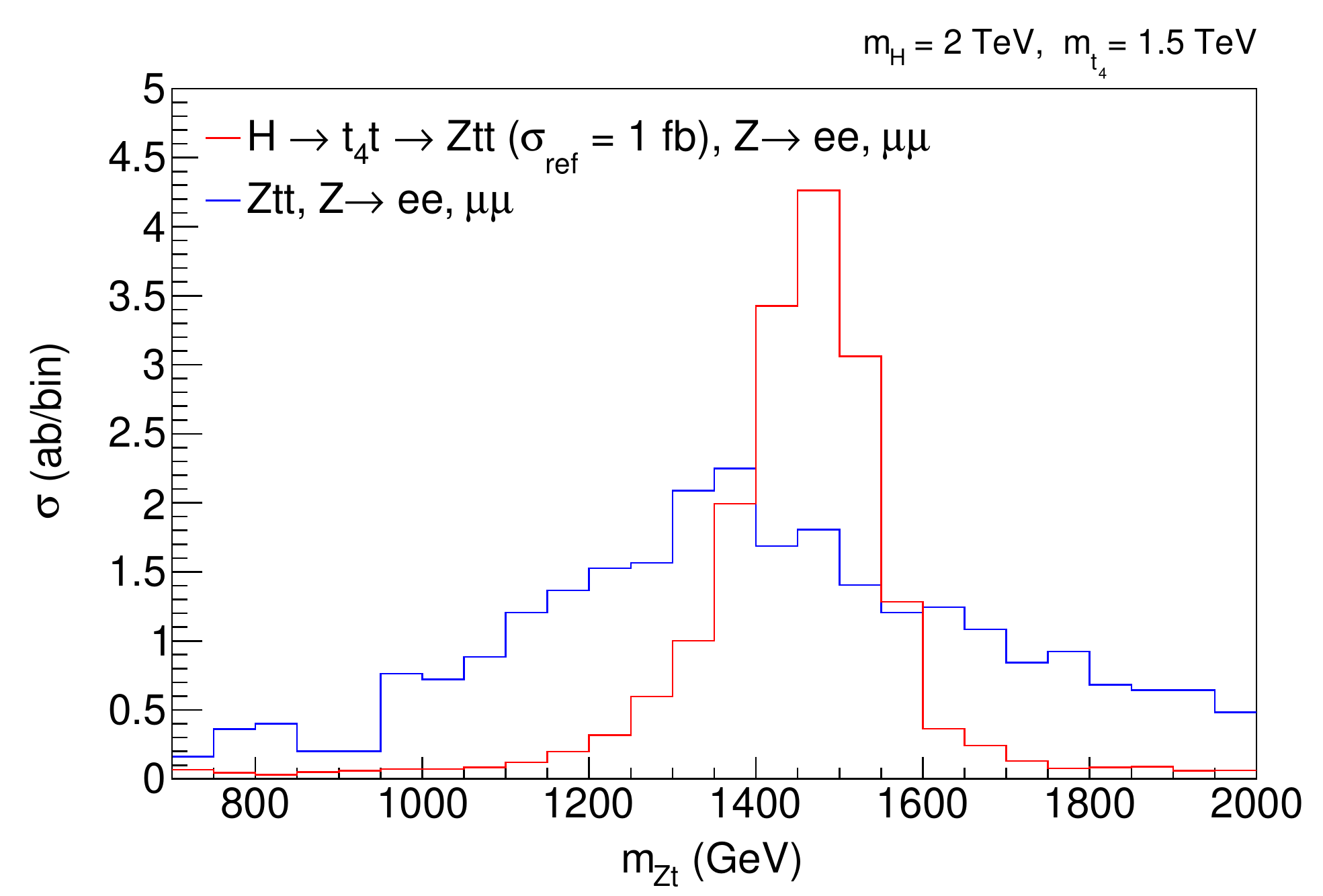}
\includegraphics[width=.49\linewidth]{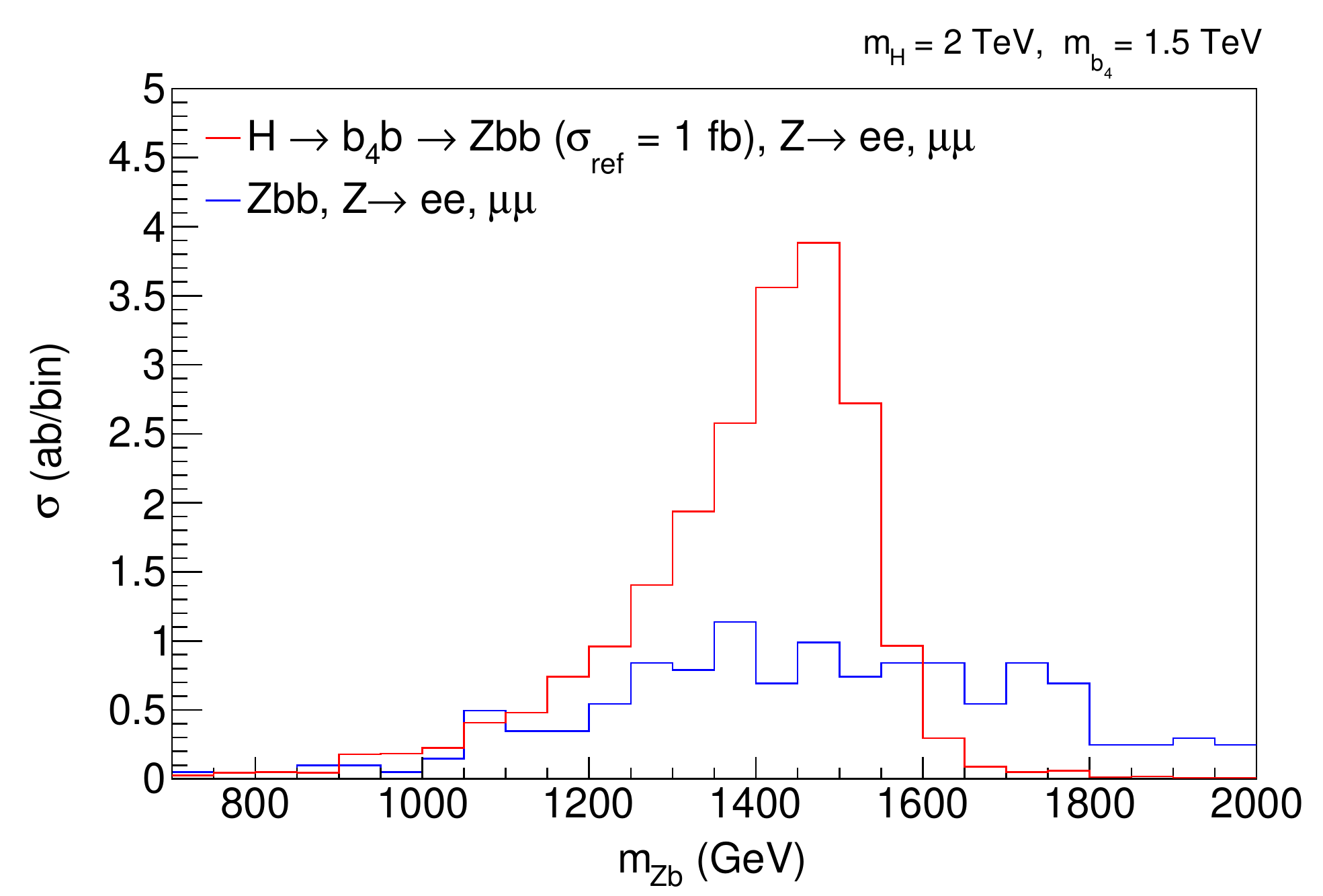}
\caption{Left panel: distribution of the invariant mass of the di-lepton pair plus the hardest fat-jet for the $H\to t_4 t\to Ztt$ signal and the corresponding $Ztt$ background, with $Z\to \ell\ell \; (\ell= e,\mu)$. Right panel: distribution of the invariant mass of the di-lepton pair plus the hardest b-jet for the $H\to b_4 b\to Zbb$ signal and the corresponding $Zbb$ background, with $Z\to \ell\ell \; (\ell= e,\mu)$. We take $m_H = 2\; {\rm TeV}$, $m_{t_4, b_4} = 1.5 \; {\rm TeV}$ and $\sigma (pp\to H\to t_4 t\to Ztt) = \sigma (pp\to H\to b_4 b\to Zbb) = 1 \; {\rm fb}$.}
\label{fig:LHCanalysis}
\end{center}
\end{figure}

Searches for $Z^\prime \to t_4 t$ by CMS have been presented in refs.~\cite{Sirunyan:2017ynj, Sirunyan:2018rfo} using 35.9 fb${}^{-1}$ of integrated luminosity. In ref.~\cite{Sirunyan:2017ynj} a search for $pp \to t_4 bj \to Z t t$ with $Z\to \ell\ell$ and hadronic top, recasted as the production and decay of a $Z^\prime$, found bounds in the range 0.06-0.13 pb. In ref.~\cite{Sirunyan:2018rfo} a dedicated search for $pp\to Z^\prime \to t_4 t \to (Wb, Zt, ht) t$ placed bounds in the range 10-100 fb in the lepton plus jets final state for $m_{t_4} <  3$ TeV and $m_{Z^\prime} < 4$ TeV. 

An important result of our analysis is that the rates for cascade decays through a bottom-like vectorlike quark ($b_4$) can be almost an order of magnitude larger than the rates for cascade decays through the top-like vectorlike quark ($t_4$). However, dedicated searches or recasted analyses for a resonance decaying to $b_4 b$ have not been performed.
Searches for the single production of bottom-like vectorlike quark in the $Wt$ and $hb$ final states by CMS have been presented in refs.~\cite{Sirunyan:2018fjh, Sirunyan:2019tib} where bounds in the 0.1-1 pb were found.
ATLAS studies of singly produced top- and bottom-like vectorlike quarks have been presented in refs.~\cite{Aaboud:2018saj, Aaboud:2018ifs, ATLAS:2018qxs} where bounds at the 0.1 pb level were found.

Estimating the reach of searches at the High Luminosity LHC (HL-LHC) with $3\; {\rm ab}^{-1}$ of integrated luminosity is challenging because of the difficulty to reproduce multi-variate analyses and to envision improvements in future analysis techniques. Nevertheless, we can obtain some conservative estimates. A naive rescaling of the analysis in ref.~\cite{Sirunyan:2018rfo} by the square root of the ratio of future to present integrated luminosities yields expected bounds at the 1-10 fb level. Future improvements in the analysis will certainly lead to lower upper bounds.  

With more data, other decay modes become competitive or even yield a stronger sensitivity. Examples include the channels $H\to t_4 t \to Z \bar t t \to \ell^+ \ell^- \bar t  t \; (\ell=e,\mu)$ with hadronic tops and $H\to b_4 b \to Z \bar bb \to \ell^+ \ell^- \bar bb \; (\ell=e,\mu)$. In this case there is no missing energy (as opposite to the analysis in ref.~\cite{Sirunyan:2018rfo}) and all invariant masses can be reconstructed. We implemented the new physics model in {\tt FeynRules}~\cite{Degrande:2011ua}. We generated parton level events with {\tt MadGraph5}~\cite{Alwall:2014hca} and used {\tt Pythia8}~\cite{Sjostrand:2006za, Sjostrand:2014zea} and {\tt Delphes}~\cite{deFavereau:2013fsa} for shower, hadronization and detector simulation. As a representative example we choose $m_H = 2 \; {\rm TeV}$ and $m_{t_4, b_4} = 1.5 \; {\rm TeV}$. In the $t_4 t$ case, we require $H_T > 1\; {\rm TeV}$, two leptons with total transverse momentum $P_T(\ell\ell) > 450 \; {\rm GeV}$, two fat-jets (as reconstructed by Delphes in its default setting) and the invariant mass of the di-lepton pair plus the hardest fat-jet $m_{j\ell\ell} \in [1.3,1.6] \; {\rm TeV}$. Similarly, in the $b_4 b$ case, we require $H_T > 1\; {\rm TeV}$, two leptons with total transverse momentum $P_T(\ell\ell) > 550 \; {\rm GeV}$, two jets out of which at least one is $b$-tagged and $m_{j\ell\ell} \in [1.2,1.6] \; {\rm TeV}$, where $j$ is the hardest jet (whether $b$-tagged or not). 

The resulting $m_{j\ell\ell}$ distributions are displayed in figure~\ref{fig:LHCanalysis} where the signals are normalized to $1\; {\rm fb}$ (prior to the $Z$ decay). From these results we estimate that the HL-LHC with an integrated luminosity of $3\; {\rm ab}^{-1}$  is sensitive to cross sections: $\sigma (pp\to H\to t_4 t\to Ztt) \gtrsim 0.19\; {\rm fb}$ and  $ \sigma (pp\to H\to b_4 b\to Zbb) \gtrsim 0.16 \; {\rm fb}$. We stress that this is a naive cut-based analysis which focuses on vectorlike quarks decaying to leptonic $Z$. We reasonably expect that, with a more sophisticated analysis and by considering/combining more decay channels, significantly smaller cross sections could be probed.   

\begin{figure}
\begin{center}
\includegraphics[width=.49\linewidth]{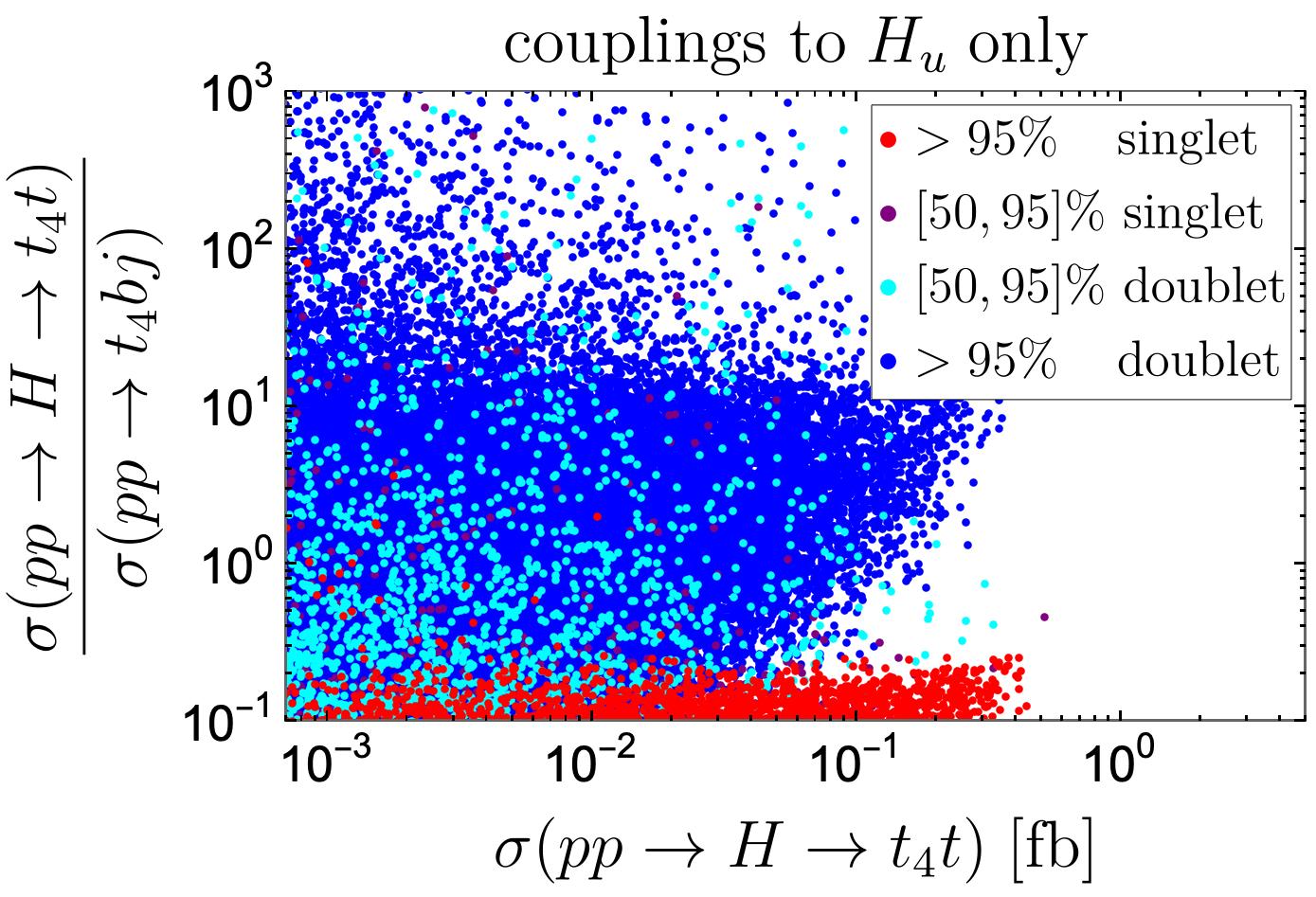}
\includegraphics[width=.49\linewidth]{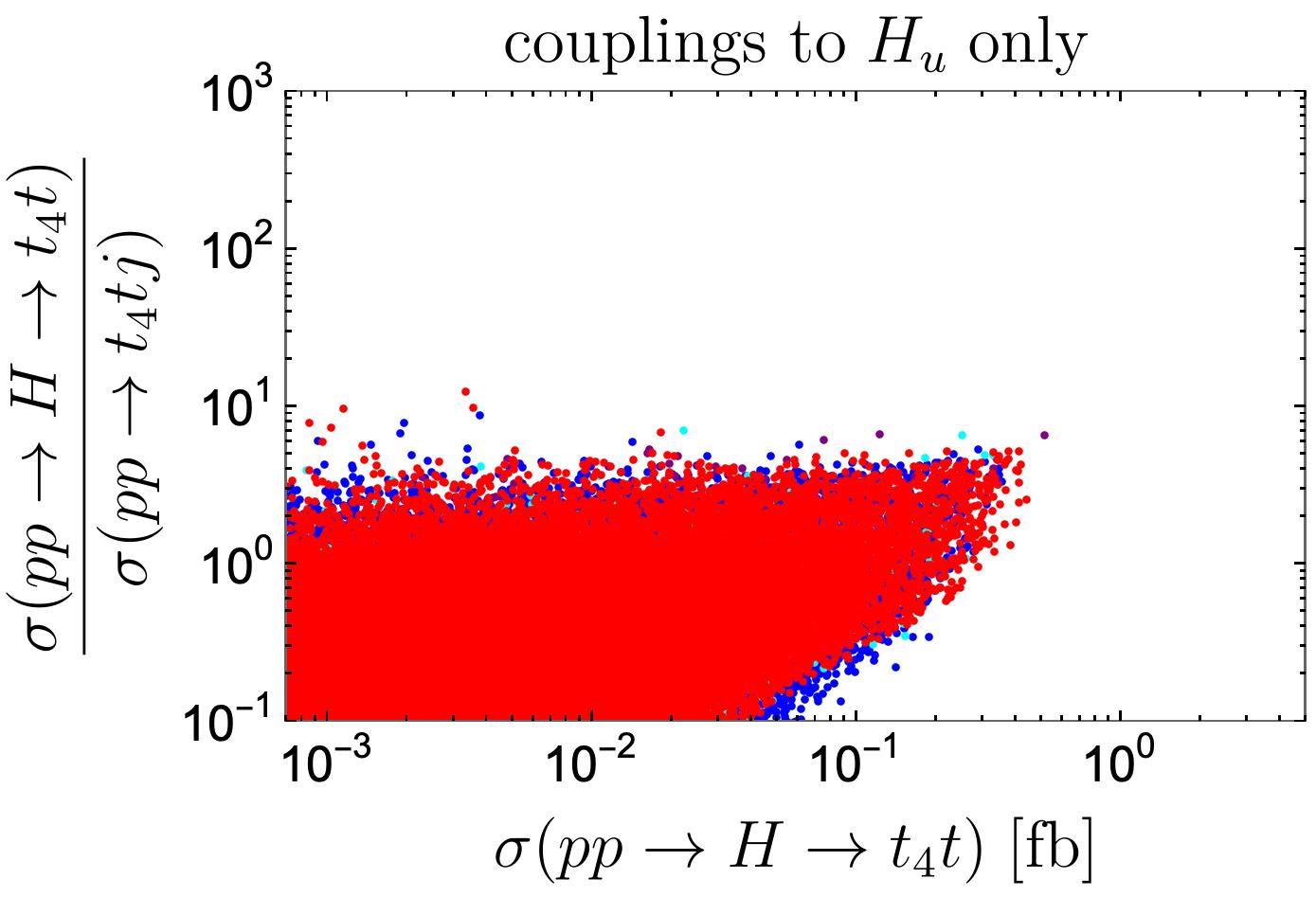}
\includegraphics[width=.49\linewidth]{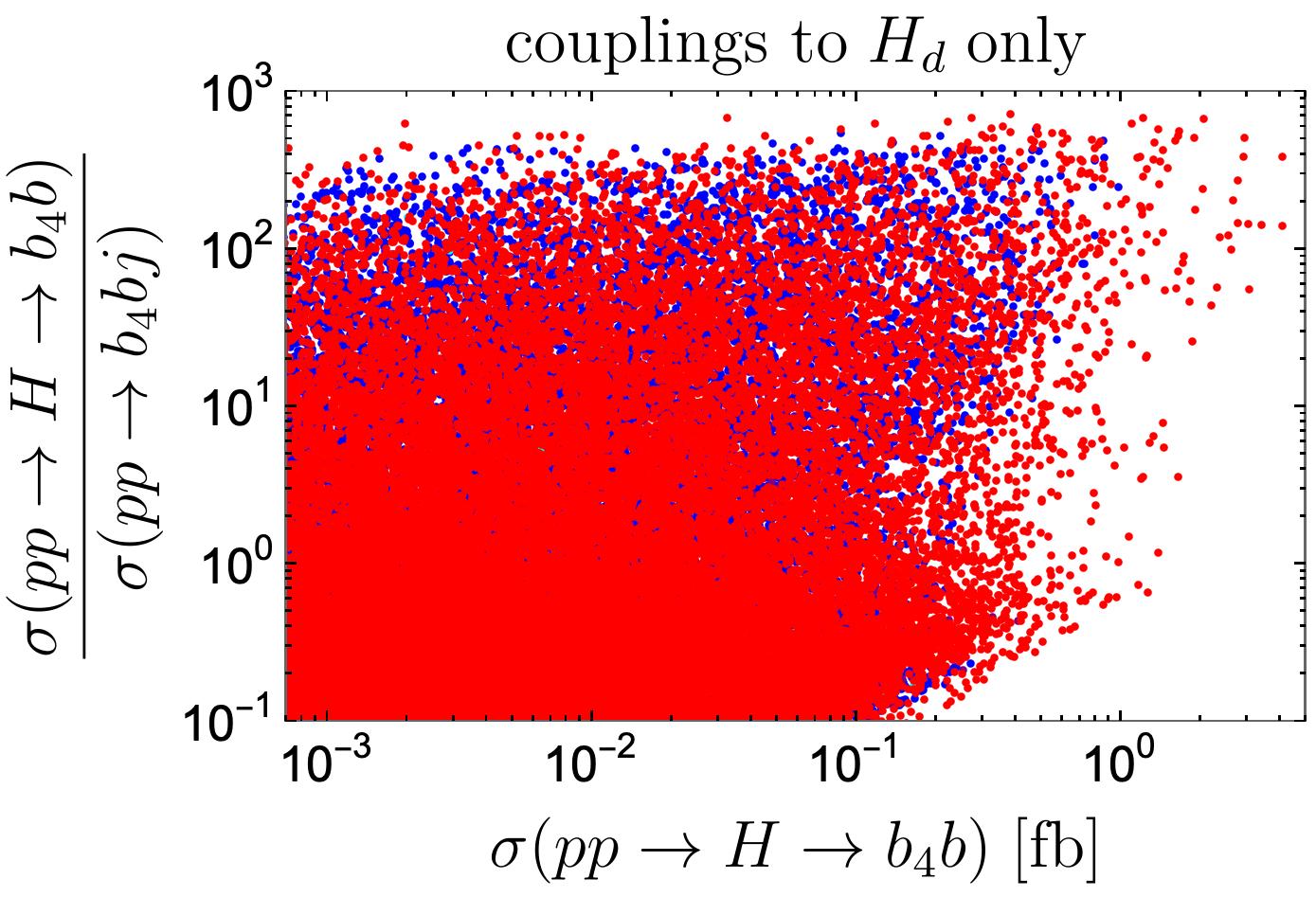}
\includegraphics[width=.49\linewidth]{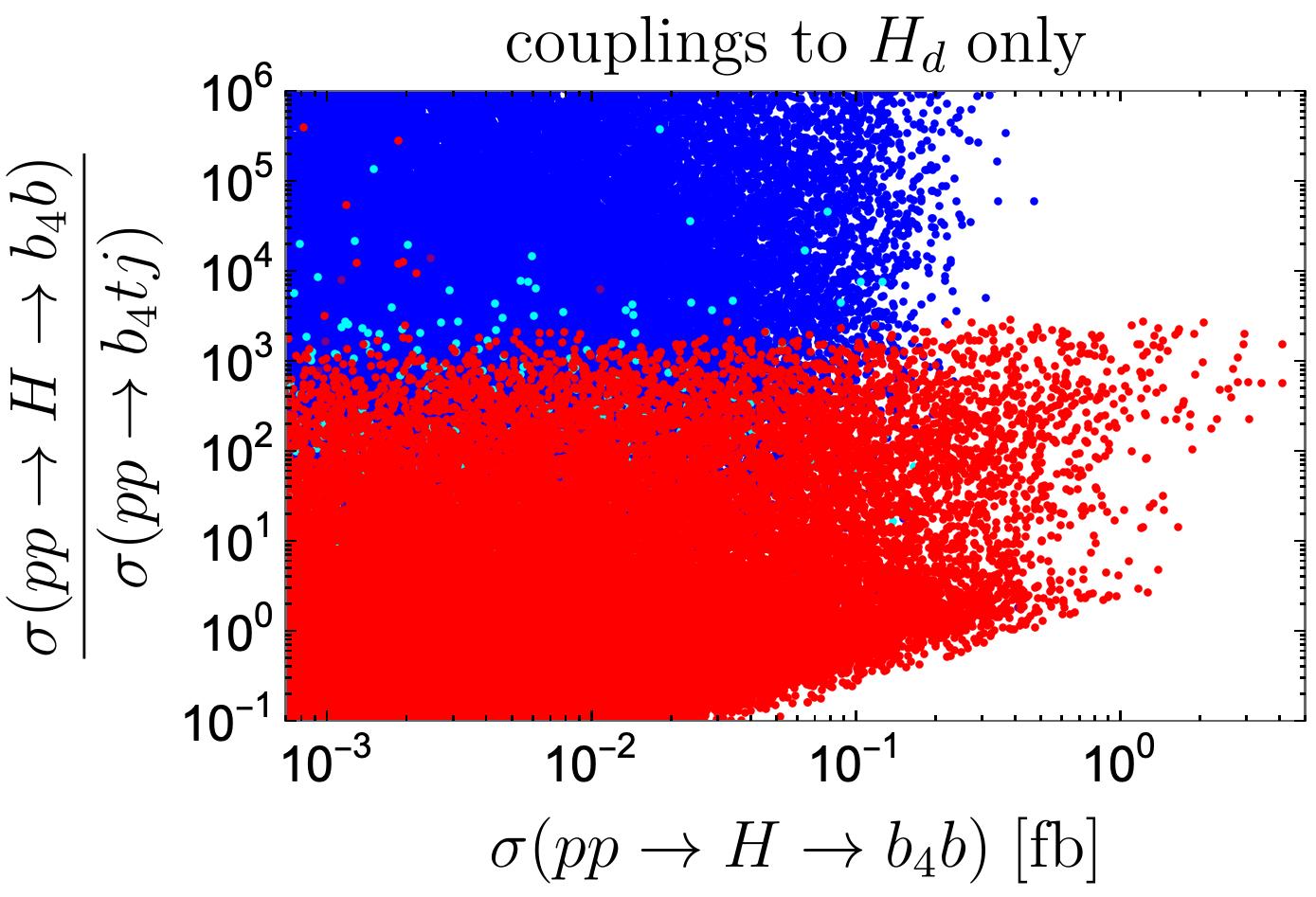}
\caption{Ratios of production rates of $t_4$ (top) and $b_4$ (bottom) in heavy Higgs cascade decays and single production modes. Results for scenarios with couplings to both $H_u$ and $H_d$ are similar.}
\label{fig:Hcomp}
\end{center}
\end{figure}

Finally, we would like to compare the production rates of vectorlike quarks in Higgs cascade decays and single production. Both mechanisms depend on the structure of Yukawa couplings  and either process can dominate. This is illustrated in figure~\ref{fig:Hcomp}, where we plot the ratios of rates for these two production mechanisms. For the single production we consider both the $W$ mediated modes, $pp\to t_4 b j$ and $pp\to b_4 t j$, and the $Z$  mediated modes, $pp\to t_4 t j$ and $pp\to b_4 b j$. Note that  $t_4$ ($b_4$) single production is typically dominated by the $W$ ($Z$) mode. We see that production rates of vectorlike quarks in Higgs cascade decays can easily be orders of magnitude larger than the single production rates. For the $b_4$, this result does not depend on its doublet/singlet nature, while for the $t_4$  it is more typical for the doublet case.\footnote{In fact, for a pure doublet $t_4$ ($b_4$) in the scenario with couplings to $H_u$ ($H_d$) only, there is no $W t_4 b$ ($W b_4 t$) vertex, implying the absence of the $W$ mediated single production. This is clearly visible in the upper-left and lower-right panels of figure~\ref{fig:Hcomp}. In scenarios with couplings to both $H_u$ and $H_d$ these $W$ couplings are not constrained to vanish and the blue points in these plots would extend to smaller values.} These findings further motivate dedicated analyses for resonances decaying into $t_4 t$ or $b_4 b$.

In contrast to Higgs cascade decays and single production mechanisms, the pair production cross section of vectorlike quarks is model independent. Projections for the reach of the HL-LHC can be found in section 6.3.2 of ref.~\cite{CidVidal:2018eel} where it is shown that vectorlike quarks with masses up to about 1350 GeV can be discovered at the 5$\sigma$ level. The corresponding 2$\sigma$ sensitivity would extend to only somewhat larger masses because of the steep decrease of the pair production cross section for increasing vectorlike quark mass. 
We see that the experimental sensitivities of the HL-LHC to vectorlike quarks in heavy Higgs cascade decays and pair production are comparable.

The cascade decays of a heavy Higgs  boson through  vectorlike quarks provide an interesting  opportunity to discover two new particles simultaneously. Not only these decay modes are sizable or can even dominate, but the usually dominant decay modes, $H\to t\bar t $ or $b\bar b$, are extremely challenging due to huge standard model backgrounds. In addition, for these decay modes, the resonant peak can be destroyed by the interference with the SM background~\cite{Jung:2015gta}.

\section{Conclusions}
\label{sec:conclusions}
We studied cascade decays of a  heavy  neutral Higgs boson through vectorlike quarks, $H\to t_4 t \to Wbt, Zt t, ht t$ and $H\to b_4 b \to Wtb, Zb b, hb b$, where $t_4$ ($b_4$) is the new up-type (down-type) quark mass eigenstate. Limiting the size of Yukawa couplings of vectorlike fields to one, in  the two Higgs doublet model type-II, we found that these decay modes can be significant or can even dominate.

We found that the  branching ratio of  $H\to t_4 t$ is larger than 10\% for $\tan\beta \in [0.5,10]$ and can reach up to $40\%$. More importantly, the branching ratio of  $H\to b_4 b$ is larger than 10\% for any $\tan\beta > 0.8$ and  this mode can dominate for $\tan\beta\in [4,18]$  reaching up to 95\%. Multiplying with the Higgs production cross section, we found that $\sigma (pp\to H\to t_4t)$ is the largest at very small $\tan\beta$ and that, while the $H\to b_4 b$ mode can dominate at medium $\tan\beta$, the $\sigma (pp\to H\to b_4b)$ can still be larger at both small and very large $\tan\beta$.

The lightest new quarks from heavy Higgs decays  further decay into SM particles through $W$, $Z$ or $h$. We studied the correlations between  the branching ratios of $H \to t_4t$ ($H \to b_4b$)  and individual branching ratios of $t_4$ ($b_4$). We presented  the maximum production rates of individual final states in cascade decays of a heavy Higgs boson as functions  of $m_H$ and $m_{t_4}$ or $m_{b_4}$. We found that, for Higgs cascade decays through a $t_4$,  rates of 0.1 fb extend up to $m_H \lesssim 2 \; \text{TeV}$ and $m_{t_4} \lesssim 1.4 \; \text{TeV}$. The rates above 0.01 fb can be achieved for $m_H \lesssim 2.8\; \text{TeV}$ or $m_{t_4} \lesssim 2 \; \text{TeV}$. For Higgs cascade decays through a $b_4$, rates of individual final states larger than 0.1 fb extend up to $m_H \lesssim  2.5\; \text{TeV}$ and $m_{b_4} \lesssim 1.8 \; \text{TeV}$ and can be even larger than 1 fb for $m_H \lesssim  1.6\; \text{TeV}$ and $m_{b_4} \lesssim 1.2 \; \text{TeV}$. The rates above 0.01 fb can be achieved for $m_H \lesssim 3.3\; \text{TeV}$ or $m_{b_4} \lesssim 2.4 \; \text{TeV}$.

The signatures of cascade decays of a heavy Higgs boson through  vectorlike quarks are almost identical to the production of any new resonance decaying to $t_4 t$ or $b_4 b$. They are also very similar  to single productions of vectorlike quarks. However the topology of cascade decays provides more handles on the final states and thus dedicated searches have a potential to considerably improve the limits found in standard single production studies. So far, only searches for a resonance decaying to $t_4 t$ have been performed. We have found that the rates for cascade decays through $b_4$ can be almost an order of magnitude larger than the rates for cascade decays through $t_4$ which motivates similar searches for a resonance decaying to $b_4 b$.

We explored the reach of decay modes in which there is no missing energy and all invariant masses can be reconstructed. As examples we considered the channels $H\to t_4 t \to Z \bar t t \to \ell^+ \ell^- \bar t  t \; (\ell=e,\mu)$ with hadronic tops and $H\to b_4 b \to Z \bar bb \to \ell^+ \ell^- \bar bb \; (\ell=e,\mu)$. We have shown that even a very simple cut based analysis allows the HL-LHC with $3\; {\rm ab}^{-1}$ of integrated luminosity to probe cross sections $\sigma (pp\to H\to t_4 t\to Ztt) \gtrsim 0.19\; {\rm fb}$ and  $ \sigma (pp\to H\to b_4 b\to Zbb) \gtrsim 0.16 \; {\rm fb}$. We expect that, with more sophisticated analyses and by considering/combining more decay channels, significantly smaller cross sections could be tested.

We have also found that the rates for these processes  can be much larger, even by orders of magnitude, than the rates for single productions of vectorlike quarks. The final states have significantly lower standard model backgrounds not only compared to single productions of vectorlike quarks but especially compared to the usually dominant decay modes of heavy Higgses, $H\to t\bar t $ or $b\bar b$, searches for which are extremely challenging. Thus, the cascade decays of a heavy Higgs  boson through  vectorlike quarks provide the best opportunities for the discovery of a new Higgs boson and vectorlike quarks.

\acknowledgments
SS thanks Bogdan Dobrescu and Eva Halkiadakis  for insightful discussion. The work of RD was supported in part by the U.S. Department of Energy under grant number {DE}-SC0010120, and the IU Institute for Advanced Study. SS is supported by the National Research Foundation of Korea (NRF-2017R1D1A1B03032076 and in partial by NRF-2018R1A4A1025334).
This work was performed in part at the Aspen Center for Physics, which is supported by National Science Foundation grant PHY-1607611. SS appreciates the hospitality of Fermi National Accelerator Laboratory and expresses a special thanks to the Mainz Institute for Theoretical Physics (MITP) of the DFG Cluster of Excellence PRISMA$^+$ (Project ID 39083149), for its hospitality and support.
This work was supported by IBS under the project code, IBS-R018-D1 (SS).

\appendix

\section{Partial decay widths of the heavy Higgs boson}
\label{app:Hwidths}
The decay widths for $H \to t_i t_j \equiv \bar t_i t_j + \bar t_j t_i$ with $i \ne j$ and $i,j=3,4,5$ are given by:
\begin{align}
\Gamma(H \to t_i t_j) &=
\frac{3 m_H}{16 \pi} \left\{ \left( \left| \lambda^H_{t_i t_j} \right|^2 +  \left| \lambda^{H\,\ast}_{t_j t_i} \right|^2  \right) \left( 1 - \frac{m_{t_i}^2 + m_{t_j}^2}{m_H^2} \right) + 4  \lambda^H_{t_i t_j}  \lambda^{H\,\ast}_{t_j t_i} \frac{m_{t_i} m_{t_j}}{m_H^2} \right\} \nonumber \\
& \hspace{1.5cm} \times  \sqrt{\lambda(1, (m_{t_i}/m_H)^2, (m_{t_j}/m_H)^2)}\,,
\label{Htitj}
\end{align}
where the couplings $\lambda^{H}_{t_i t_j}$ are given in Eq.~(A.48) of ref.~\cite{Dermisek:2019vkc} and
$\lambda(x,y,z) = x^2 + y^2 + z^2 - 2 xy - 2yz - 2zx$.

The decay widths for $H\to \bar t_i t_i$ with $i=3,4,5$ are given by:
\begin{align}
\Gamma(H \to t_i t_i) &= \frac{3 m_H}{16 \pi}|\lambda^H_{t_i t_i}|^2 \left( 1 - \frac{4 m_{t_i}^2 }{m_H^2} \right)^{3/2}\,.
\label{Htiti}
\end{align}
Note that the corresponding partial widths into $b_i$ $(i=3,4,5)$ can be obtained from Eqs.~(\ref{Htitj}) and (\ref{Htiti}) by replacing $t_{i,j} \to b_{i,j}$. For the $H\to \bar t t$ case we include the NLO correction factor $\left[1+\frac43 \frac{\alpha_s}{\pi} \Delta^t_H (\beta_t)\right]$ where $\beta_t = 1 - 4 m_t^2 / m_H^2$ and $\Delta^t_H (\beta_t)$ is given in Eq.~(2.14) of ref.~\cite{Djouadi:2005gi}.

The decay width for $H \to \bar b b$ in the $m_b \ll m_H$ limit is given by:
\begin{align}
  \Gamma(H \to \bar b b) &= \frac{3 m_H}{16 \pi} \left| \lambda^H_{bb} \right|^2 \left[ 1 + \Delta_{qq} + \Delta_H^2 \right],
\end{align}
where the coupling $\lambda^H_{bb}$ should be evaluated at a scale $O(m_H)$ and the N${}^3$LO correction factors, $\Delta_{qq}$ and $\Delta_H^2$, are presented in Eqs.~(2.11) and (2.12) of ref.~\cite{Djouadi:2005gi}.

The decay width of $H \to gg$ is given by:
\begin{align}
\Gamma(H \to g g) &= \frac{G_F \alpha_s^2 m_H^3}{36 \sqrt{2} \pi^3} \cdot \frac{9}{16} \, \left| -  A_{1/2}^t \cot\beta + A_{1/2}^b \tan\beta + \frac{\lambda^H_{t_4 t_4}  v}{m_{t_4}} A_{1/2}^{t_4} +  \frac{\lambda^H_{t_5 t_5} v}{m_{t_5}} A_{1/2}^{t_5} \right. \nonumber  \\
&\hspace{3.5cm} \left. +  \frac{\lambda^H_{b_4 b_4} v}{m_{b_4}} A_{1/2}^{b_4} +  \frac{\lambda^H_{b_5 b_5} v}{m_{b_5}} A_{1/2}^{b_5}   \right|^2~,
\end{align}
where $A_{1/2}^{q} = 2[\tau_{q} + (\tau_{q} - 1) f(\tau_{q})]/\tau_{q}^2$ with  $\tau_q = m_H^2 / 4 m_q^2$ and
\begin{align}
f(\tau) =
\begin{cases}
  \arcsin^2 \sqrt{\tau}, & \tau \le 1 \cr
-\frac{1}{4} \left[ \log\frac{1+\sqrt{1-\tau^{-1}}}{1-\sqrt{1-\tau^{-1}}}-i \pi \right]^2, & \tau > 1 \cr
\end{cases} \; .
\end{align}
Note that the  contributions of vectorlike quarks have been included following the prescription discussed in refs.~\cite{Cacciapaglia:2009ky,Cacciapaglia:2011fx}.

The decay width for $H \to \gamma \gamma$ can be obtained from Eq.~(2.23) of ref.~\cite{Djouadi:2005gj} supplemented by the presecription discussed in refs.~\cite{Cacciapaglia:2009ky,Cacciapaglia:2011fx} and is given by:
\begin{align}
\Gamma(H \to \gamma \gamma) &= \frac{9 G_F \alpha^2 m_H^3}{128 \sqrt{2} \pi^3}  \left| \left( \frac23 \right)^2 (-\cot\beta) \cdot A_{1/2} (\tau_t) + \left( \frac13 \right)^2 \tan\beta \cdot A_{1/2} (\tau_b) \right. \nonumber \\
&\hspace{1.8cm} + \left( \frac23 \right)^2 \cdot \frac{\lambda^H_{t_4 t_4} v}{m_{t_4}}  \cdot  A_{1/2} (\tau_{t_4}) + \left( \frac23 \right)^2 \cdot \frac{\lambda^H_{t_5 t_5} v}{m_{t_5}} \cdot  A_{1/2} (\tau_{t_5}) \nonumber \\
&\hspace{1.5cm} \left. +  \left( \frac13 \right)^2 \cdot \frac{\lambda^H_{b_4 b_4} v}{m_{b_4}} \cdot  A_{1/2} (\tau_{b_4}) + \left( \frac13 \right)^2 \cdot \frac{\lambda^H_{b_5 b_5} v}{m_{b_5}} \cdot  A_{1/2} (\tau_{b_5}) \right|^2.
\end{align}

Finally, the decay width for $H \to h h$, where $h$ is the SM-like Higgs boson, is given by:
\begin{align}
\Gamma(H \to h h) &=  \frac{9 G_F}{16 \pi \sqrt{2}} \frac{M_Z^4}{m_H} \sin^2 (2\beta) \cos^2 (2\beta) \sqrt{1 - 4 \frac{m_h^2}{m_H^2}} \; .
\label{eq:Hhh}
\end{align}

  \end{document}